\documentclass[journal,draftcls,onecolumn,12pt,twoside]{IEEEtranTCOM}
\usepackage{amsmath}
\usepackage{amsfonts}
\usepackage{cite}
\usepackage{amssymb}
\usepackage{mathrsfs}
\usepackage{graphicx}
\usepackage[usenames,dvipsnames]{color}
\usepackage{epstopdf}
\usepackage[all]{xy}
\usepackage[]{mcode}
\usepackage{supertabular}
\usepackage{subcaption}
\usepackage{array}

\usepackage{algorithmic}
\usepackage{algorithm}

\providecommand{\U}[1]{\protect\rule{.1in}{.1in}}
\pdfoutput=1
\providecommand{\U}[1]{\protect\rule{.1in}{.1in}}

\newcommand{\qed}{\nobreak \ifvmode \relax \else
      \ifdim\lastskip<1.5em \hskip-\lastskip
      \hskip1.5em plus0em minus0.5em \fi \nobreak
      \vrule height0.75em width0.5em depth0.25em\fi}

\begin{document}

\title{On Telecommunication Service Imbalance and Infrastructure Resource Deployment}
\author{Chuanting Zhang, \textit{Member, IEEE}, Shuping Dang, \textit{Member, IEEE}, Basem Shihada, \textit{Senior Member, IEEE}, and Mohamed-Slim Alouini, \textit{Fellow, IEEE}
  \thanks{The authors are with Computer, Electrical and Mathematical Science and Engineering Division, King Abdullah University of Science and Technology (KAUST), 
Thuwal 23955-6900, Saudi Arabia (e-mail: \{chuanting.zhang, shuping.dang, basem.shihada, slim.alouini\}@kaust.edu.sa).}}

\maketitle
\date{}

\begin{abstract}
The digital divide restricting the access of people living in developing areas to the benefits of modern information and communications technologies has become a major challenge and research focus. To well understand and finally bridge the digital divide, we first need to discover a proper measure to characterize and quantify the telecommunication service imbalance. In this regard, we propose a fine-grained and easy-to-compute imbalance index, aiming to quantitatively link the relation among telecommunication service imbalance, telecommunication infrastructure, and demographic distribution. The mathematically elegant and generic form of the imbalance index allows consistent analyses for heterogeneous scenarios and can be easily tailored to incorporate different telecommunication policies and application scenarios. Based on this index, we also propose an  infrastructure resource deployment strategy by minimizing the average imbalance index of any geographical segment. Experimental results verify the effectiveness of the proposed imbalance index by showing a high degree of correlation to existing congeneric but coarse-grained measures and the superiority of the infrastructure resource deployment strategy.
\end{abstract}


\section{Introduction}
\IEEEPARstart{T}{here} has been a consensus reached in the communications research community that sixth generation (6G) communications should focus on the solution to the digital divide that restricts current 3.6 billion of the world's population from enjoying the benefits of latest information and communications technologies (ICTs) \cite{Yaacoub2020}. Accordingly, in the context of 6G communications, the major ICT research roadmap is being redirected from boosting the capacity-related metrics in populous areas to democratizing the benefits of ICTs over the globe \cite{dang2020should}. It is widely believed that the digital inclusion and the accessibility to the benefits of ICTs in underdeveloped areas will be the key impetus for economic growth and progress in education, health care, and the administration of public affairs \cite{philbeck2017connecting}. Because of these incentives and the universal faith in creating a connected and fair world, the research community and governing bodies have jointly set a prospective timeline for gradually bridging the digital divide \cite{broadband2020}.

Unfortunately, although the campaign for fair access to the ICT benefits has been launched over the past few years, the actual progress is barely satisfactory. This is because the causes of the digital divide are highly complex and coupled with both technological and non-technological factors \cite{Khaturia2020}. Furthermore, a comprehensive and accurate measure to quantify the telecommunication service imbalance is still lacking, which results in a challenge for quantitative evaluations and analyses. Though the GSM Association (GSMA) released the mobile connectivity index in 2014 \cite{GSMA_2020}, and Facebook invented the inclusive internet index in 2015 \cite{Inclusive_2020}, these indexes provide only country-wise statistics and cannot reflect fine-grained situations corresponding to geographical segments. The coarse-grained nature of these two indexes restricts their general-purpose applications.

Thus, to expedite the democratizing procedure for ICT benefits, a fine-grained measure of the telecommunication service imbalance is in dire need. In this regard, we propose a fine-grained and easy-to-compute imbalance index in this letter to characterize and quantify such imbalance. The proposed imbalance index has several analytical and computing advantages, and there are two free parameters in the proposed index that can be customized to incorporate different telecommunication policies and application scenarios. On the basis of the proposed imbalance index, we also propose an infrastructure resource deployment strategy by minimizing the average imbalance of a given segment. We carry out a series of experiments to verify the effectiveness of the proposed imbalance index and its consistency with other congeneric but coarse-grained measures. The proposed imbalance index can effectively help to identify the digital divides in different levels and pave the way for bridging them in the near future. 



%

\section{Design of the Imbalance Index}\label{dii}

\begin{figure*}
     \centering
     \begin{subfigure}[b]{0.45\textwidth}
         \centering
         \includegraphics[width=\textwidth]{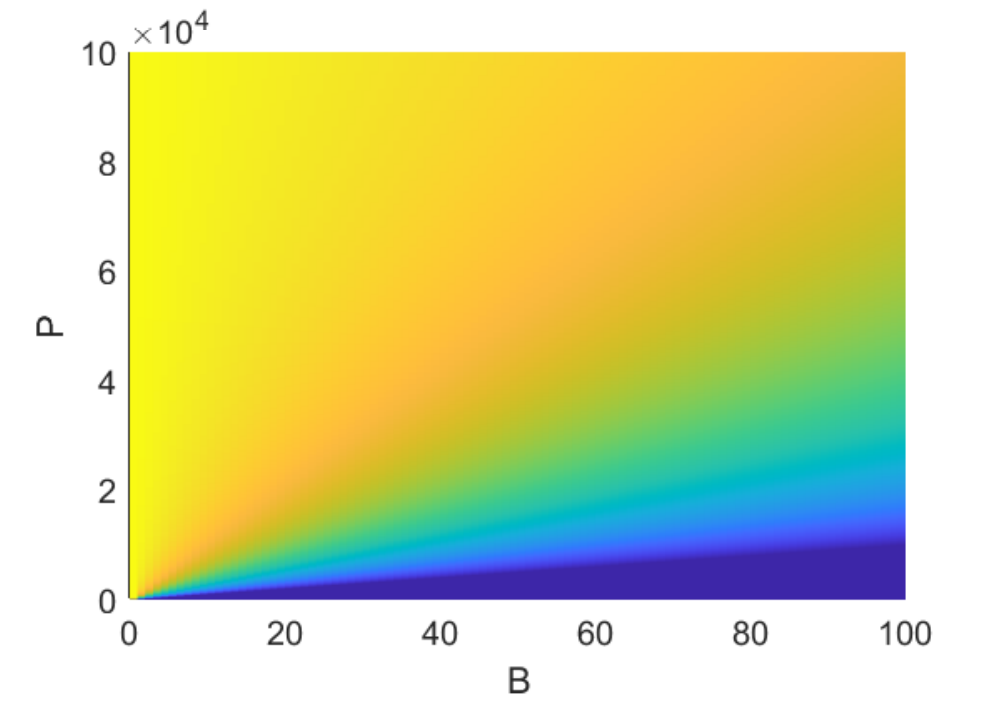}
         \caption{$\rho=100$, $\alpha=1$, $\beta=1$}
         \label{case1}
     \end{subfigure}
     \hfill
     \begin{subfigure}[b]{0.45\textwidth}
         \centering
         \includegraphics[width=\textwidth]{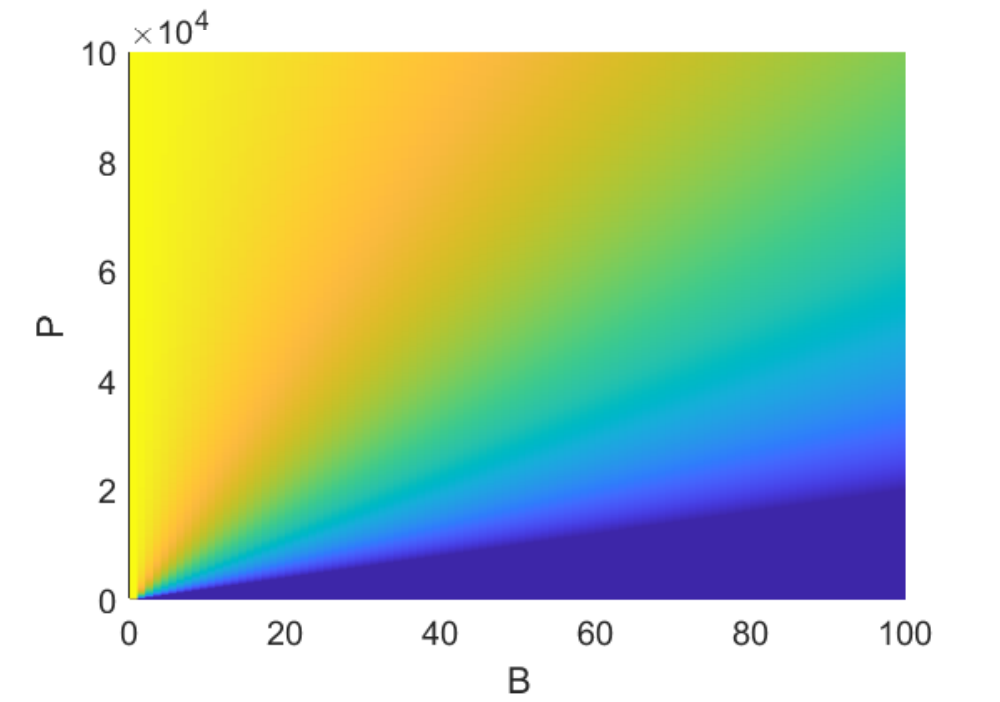}
         \caption{$\rho=200$, $\alpha=1$, $\beta=1$}
         \label{case2}
     \end{subfigure}
     
     \begin{subfigure}[b]{0.45\textwidth}
         \centering
         \includegraphics[width=\textwidth]{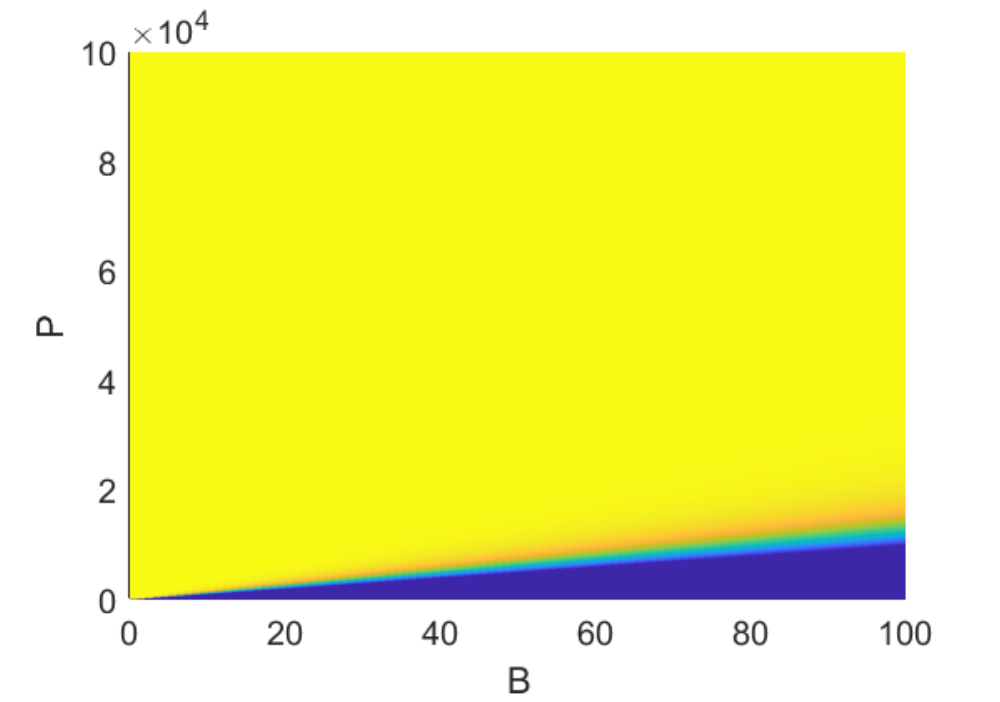}
         \caption{$\rho=100$, $\alpha=5$, $\beta=1$}
         \label{case3}
     \end{subfigure}
     \hfill
     \begin{subfigure}[b]{0.45\textwidth}
         \centering
         \includegraphics[width=\textwidth]{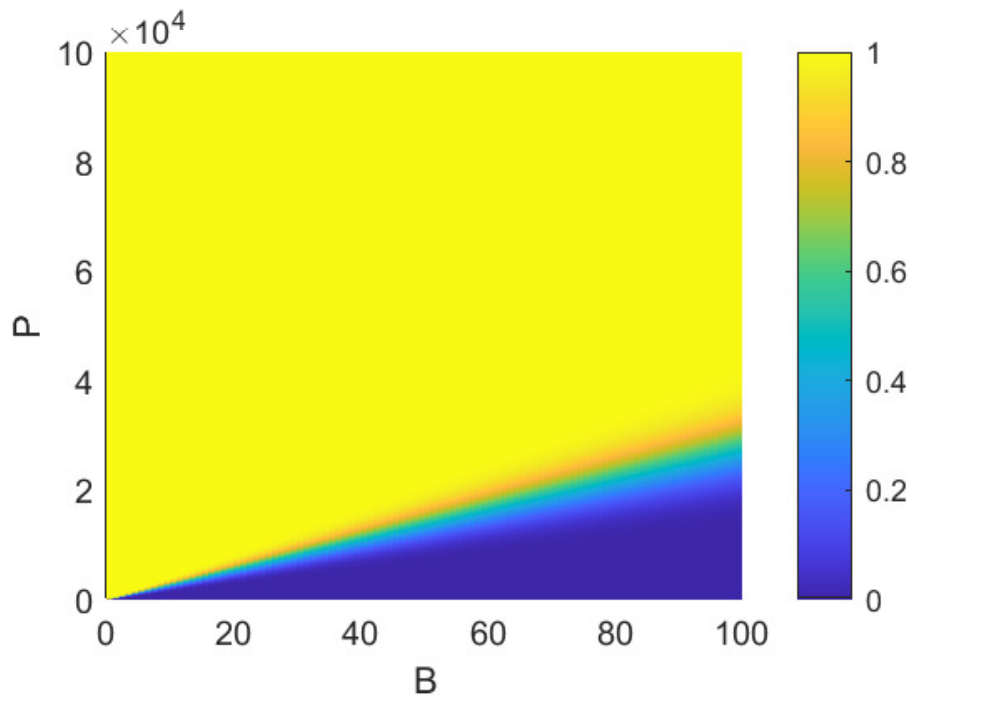}
         \caption{$\rho=100$, $\alpha=1$, $\beta=5$}
         \label{case4}
     \end{subfigure}
        \caption{Imbalance index versus population and the number of BSs in a single granule when $\rho$, $\alpha$, and $\beta$ change.}
        \label{index_changes}
\end{figure*}


Based on the appraisals of the existing works and aforementioned coarse-grained measures, we can summarize a series of ideal features for a well-designed imbalance index for telecommunication services, which should be kept as the design principles:
\begin{itemize}
\item The imbalance index should be a bounded measure between 0 and 1, corresponding to the completely fair and unfair cases, respectively;
\item The imbalance index should be a monotone non-decreasing function with respect to population;
\item The imbalance index should be a monotone non-increasing function with respect to the amount of launched telecommunication infrastructures;
\item The imbalance index should be a continuous, smooth, differentiable, and integrable function;
\item The imbalance index should be able to be customized to incorporate various requirements imposed by different telecommunication policies and application scenarios;
\item The imbalance index should be fine-grained and can geographically reflect the imbalance circumstance in detail;
\item The imbalance index should be easy-to-compute and does not involve an indefinite number of steps and special functions for computation;
\item The imbalance index should keep the analytical consistency with other existing coarse-grained measures.
\end{itemize}

Taking all the ideal features listed above into consideration, we were enlightened by the designs of logistic and Prelec functions, which are widely applied in data science and economics, respectively, and propose the following imbalance index for characterizing telecommunication service imbalance of a specific geographical segment (a.k.a. a granule). For a granule whose area is far larger than the coverage of a single base station (BS), we define
\begin{equation}\label{inequalityindexsinglegreq}
\begin{split}
\lambda\left(P,B,\boldsymbol{\rho}(B)\right)=\begin{cases}2 \left[1+ \exp\left(-\alpha\left(\log\left(\frac{P}{\sum_{i=1}^{B}\rho_i}\right)\right)^\beta\right)\right]^{-1}-1,~P>\sum_{i=1}^{B}\rho_i\\
0,~~~~~~~~~~~~~~~~~~~~~~~~~~~~~~~~~~~~~~~~~~~~~~~~~~~P\leq \sum_{i=1}^{B}\rho_i
\end{cases}
\end{split}
\end{equation}
to measure the telecommunication service imbalance, where $P$ and $B$ are the population and the number of BSs within the specific geographical segment; $\boldsymbol{\rho}(B)=[\rho_1,\rho_2,\dots,\rho_B]$ is the specification vector, and $\rho_i$ is the number of users that can be supported by the $i$th BS within this granule; $\alpha>0$ and $\beta>0$ are two free parameters that can be customized to incorporate various requirements imposed by different telecommunication policies and application scenarios. 

Note that, $\sum_{i=1}^{B}\rho_i$ is the total number of users that can be supported by $B$ BSs in this segment and can be reduced to $\rho B$ if all BSs are homogeneous, i.e., $\rho_1=\rho_2=\dots=\rho_B=\rho$. It should also be emphasized that the calculation of the imbalance index is only related to the local infrastructure and demographic information within a granule \textit{per se} but irrelevant to the global information encompassing other granules. As a result, the proposed imbalance index is advantageous over those indexes requiring global information, e.g., Gini index. Meanwhile, it is obvious that all aforementioned ideal features for a well-designed imbalance index are satisfied or partially satisfied (the analytical consistency with representative coarse-grained measures will be shown in Section \ref{erd}).

To reveal the impacts of the arguments and parameters, we assume a homogeneous application scenario with $\rho_1=\rho_2=\dots=\rho_B=\rho$ and plot the imbalance index for a single granule given different parameters in Fig.~\ref{index_changes}. From this figure, it is clear that updating BSs so as to have a larger $\rho$ is an effective way to improve telecommunication service imbalance. Furthermore, both free parameters $\alpha$ and $\beta$ can be jointly customized to vary the imbalance index and thereby reflect the doctrine of telecommunication policies. e.g., to be individualist or collectivist. Besides, $\alpha$ and $\beta$ can also be designed in a data-oriented manner so as to produce the maximum of correlation between $\lambda\left(P,B,\boldsymbol{\rho}(B)\right)$ and a certain coarse-grained measure.

For a region consisting of $M$ granules, represented by set $\mathcal{R}(M)$, we can introduce the average imbalance index as follows to reflect the telecommunication service imbalance over the entire region
\begin{equation}\label{inequalityindexeqoriginal}
\bar{\lambda}=\left.{\sum_{m\in\mathcal{R}(M)}A_m\lambda\left(P_m,B_m,\boldsymbol{\rho}_m(B_m)\right)}\middle /{\sum_{m\in\mathcal{R}(M)}A_m}\right.,
\end{equation} 
where $A_m$ is the area of the $m$th granule. The average imbalance index given in (\ref{inequalityindexeqoriginal}) can be reduced to $\bar{\lambda}=\frac{1}{M}\underset{m\in\mathcal{R}(M)}{\sum}\lambda\left(P_m,B_m,\boldsymbol{\rho}_m(B_m)\right)$, if the region is uniformly gridded with $M$ granules.

\section{Infrastructure Resource Deployment}\label{ords}
With $\bar{\lambda}$, we can formulate a simplistic telecommunication infrastructure allocation/BS placement problem that aims to provide fair access to telecommunication services\footnote{Here, we only demonstrate a simplistic application of the imbalance index. However, more realistic application scenarios can be optimized and analyzed in a similar manner by imposing extra optimization constraints.}:
\begin{equation}\label{opteq}
\begin{split}
&\mathrm{Minimize}:~\bar{\lambda}\\
&\mathrm{s.t.,}~\sum_{m\in\mathcal{R}(M)}B_m\leq B_{\max}\\
&~~~~~~~B_m\geq 0~\mathrm{and}~B_m\in\mathbb{N},~\forall~m\in\mathcal{R}(M)\\
&~~~~~~~\rho_{m,i}=\rho,~\forall~m\in\mathcal{R}(M)~\mathrm{and}~1\leq i\leq B_m,
\end{split}
\end{equation}
where $B_{\max}$ is the maximum number of BSs that are available to be deployed over the entire region, which is directly restricted by implementation cost. This formulated problem can provide a new dimension and guidelines for telecommunication planning and construction.

From (\ref{inequalityindexsinglegreq}) and (\ref{inequalityindexeqoriginal}), we know that $\bar{\lambda}$ is a quasi-convex function with respect to $\{B_m\}$ through monotonicity analysis. Therefore, for such a simplistic case, we can first relax the condition that $B_m$ must be an integer and employ the Lagrangian method to derive a sub-optimal allocation scheme of $B_{\max}$ BSs over the entire region consisting of $M$ granules. In particular, we construct the Lagrangian function by
\begin{equation}
L(\{B_m\},\nu)=\bar{\lambda}+\nu\left( B_{\max}-\sum_{m\in\mathcal{R}(M)}B_m\right)+\sum_{m\in\mathcal{R}(M)}\upsilon_mB_m,
\end{equation}
where $\nu$ and $\{\upsilon_m\}$ are the Lagrange multipliers. Subsequently, because when $P_m>\rho B_m$, $\forall~m\in\mathcal{R}(M)$, the optimization problem formulated in (\ref{opteq}) becomes a standard nonlinear programming problem by relaxing $B_m\in\mathbb{N}$, we can derive the following Karush-Kuhn-Tucker (KKT) conditions:
\begin{equation}\label{sakdjlkdeqda51}
\begin{cases}
\frac{\partial L}{\partial B_m}=\frac{-A_m\alpha\beta\left(\log\left(\frac{P_m}{\rho B_m}\right)\right)^{\beta-1}}{\left({\underset{n\in\mathcal{R}(M)}{\sum}A_n}\right)B_m\left[1+\cosh\left(\alpha\left(\log\left(\frac{P_m}{\rho B_m}\right)\right)^{\beta}\right)\right]}-\nu+\upsilon_m\\
~~~~~~=0,~\forall m\in\mathcal{R}(M)\\
\frac{\partial L}{\partial \nu}= B_{\max}-\underset{m\in\mathcal{R}(M)}{\sum}B_m=0\\
\upsilon_m B_m=0,~\forall m\in\mathcal{R}(M)
\end{cases}
\end{equation}
However, since solving process of (\ref{sakdjlkdeqda51}) involves transcendental equations, to the best of authors' knowledge, there is no closed-form
solution for arbitrary $\alpha$ and $\beta$. For a special case when $\alpha=\beta=1$, we can generally express the standard water-filling solution of $B_m$ as
\begin{equation}
B_m^*=\left[\sqrt{\frac{2A_mP_m}{(\upsilon_m-\nu)\rho\underset{n\in\mathcal{R}(M)}{\sum}A_n}}-\frac{P_m}{\rho}\right]^+,
\end{equation}
where $[\cdot]^+=\max\{0,\cdot\}$; $\nu$ and $\{\upsilon_m\}$ should satisfy
\begin{equation}
\begin{cases}
B_{\max}-\underset{n\in\mathcal{R}(M)}{\sum}\left(\sqrt{\frac{2A_mP_m}{(\upsilon_m-\nu)\rho\underset{n\in\mathcal{R}(M)}{\sum}A_n}}-\frac{P_m}{\rho}\right)=0\\
\upsilon_m\left[\sqrt{\frac{2A_mP_m}{(\upsilon_m-\nu)\rho\underset{n\in\mathcal{R}(M)}{\sum}A_n}}-\frac{P_m}{\rho}\right]^+=0
\end{cases},
\end{equation}
respectively. A general and closed-form expression of $B_m^*$ independent from $\nu$ and $\upsilon_m$ does not exist, since the solution is coupled with the solutions of other $M-1$ granules. For a given demographic distribution over $\mathcal{R}(M)$, iterative approaching methods, e.g., the water-filing algorithm, can be utilized to obtain numerical solutions of $\{B_m^*\}$ \cite{Bansal2010}.

\begin{figure}
     \centering
     \begin{subfigure}[b]{0.4\textwidth}
         \centering
         \includegraphics[width=\textwidth]{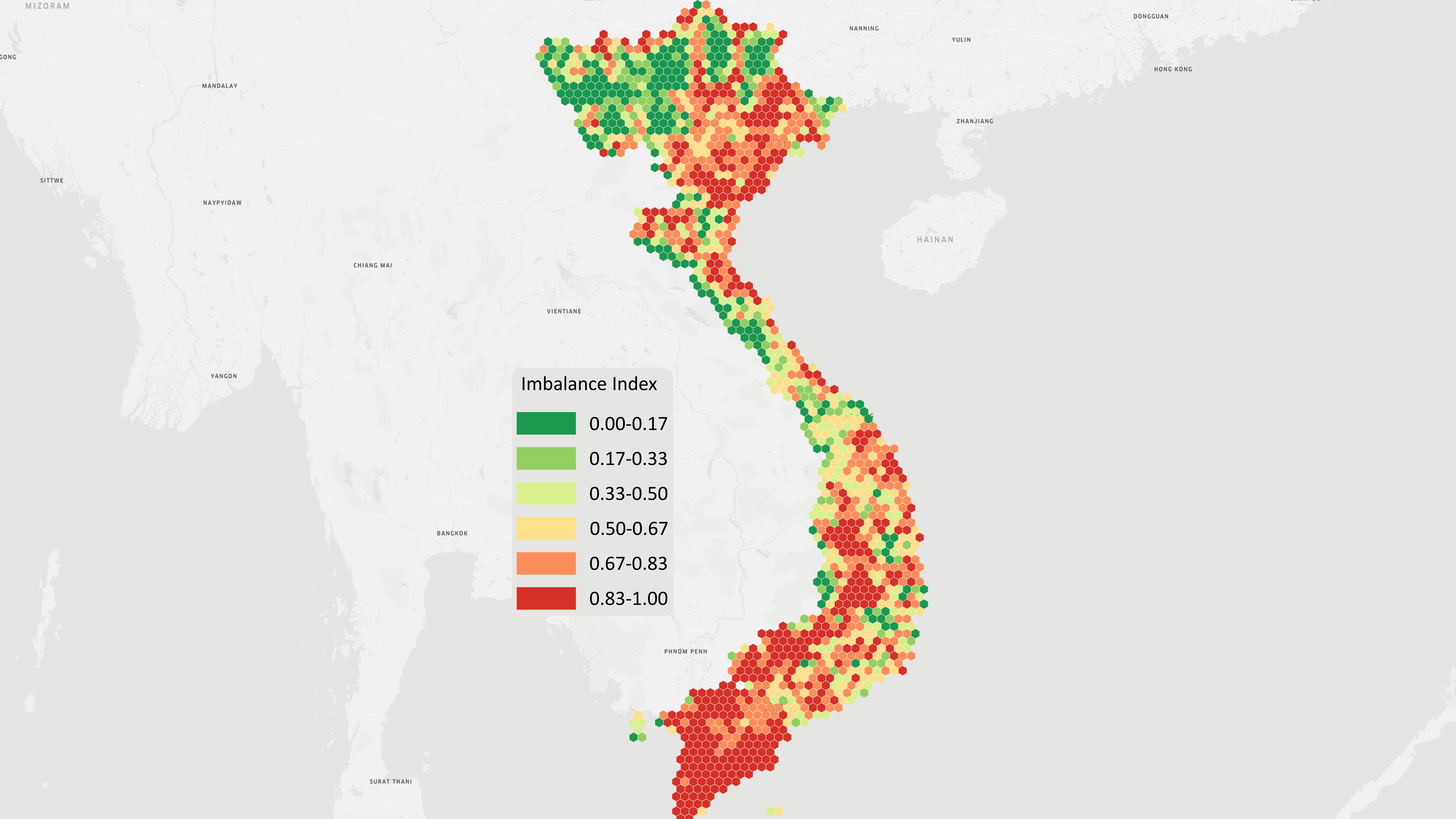}
         \caption{}
         \label{vnm_grid}
     \end{subfigure}
     \hfill
     \begin{subfigure}[b]{0.4\textwidth}
         \centering
         \includegraphics[width=\textwidth]{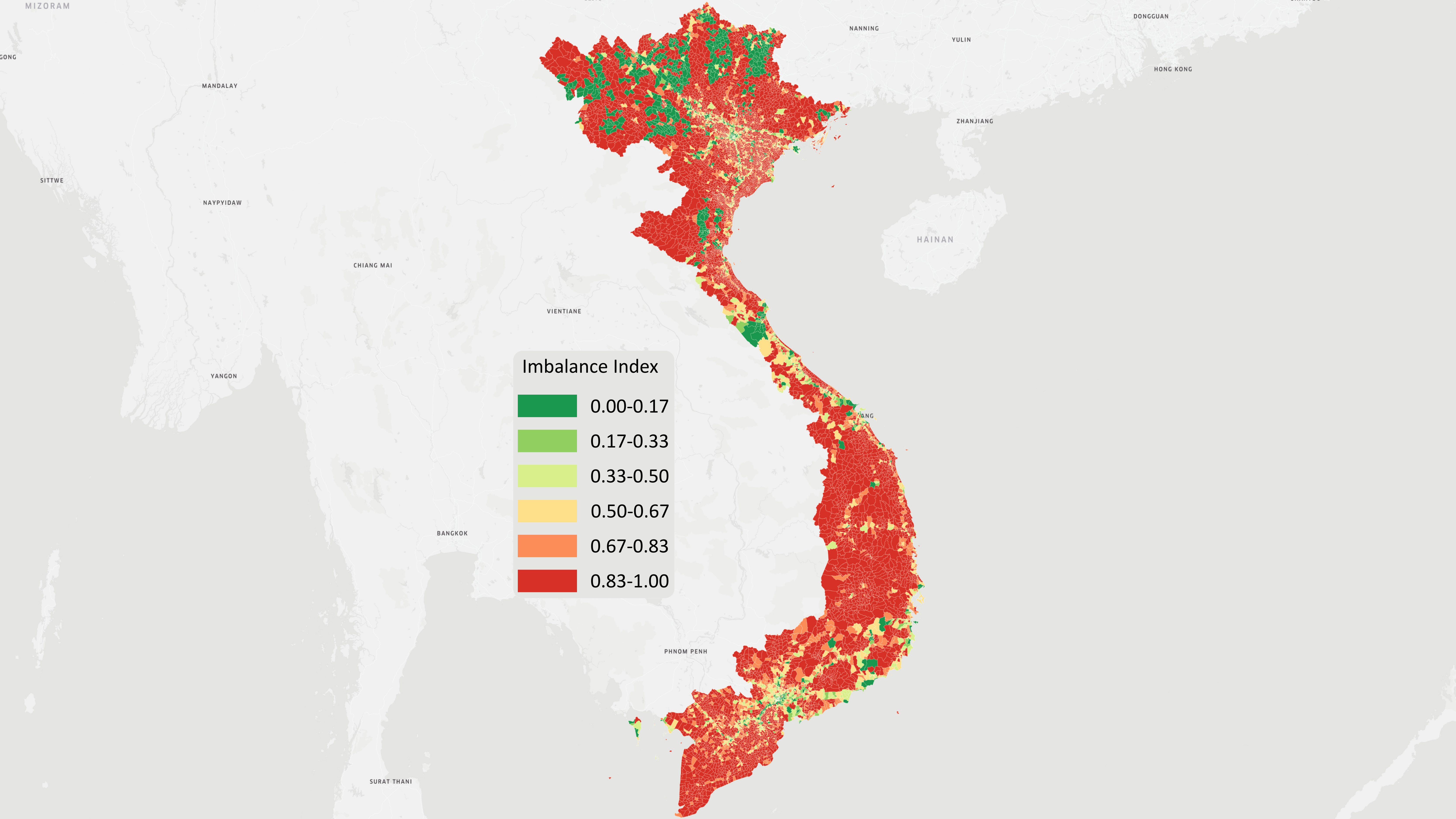}
         \caption{}
         \label{vnm_division}
     \end{subfigure}
     \caption{Telecommunication service imbalance index visualization for Vietnam. (a) Grid-wise imbalance index visualization. (b) Administrative division-wise imbalance index visualization.}
        \label{index_vnm}
\end{figure}

\begin{figure}
     \centering
     \begin{subfigure}[b]{0.45\textwidth}
         \centering
         \includegraphics[width=\textwidth]{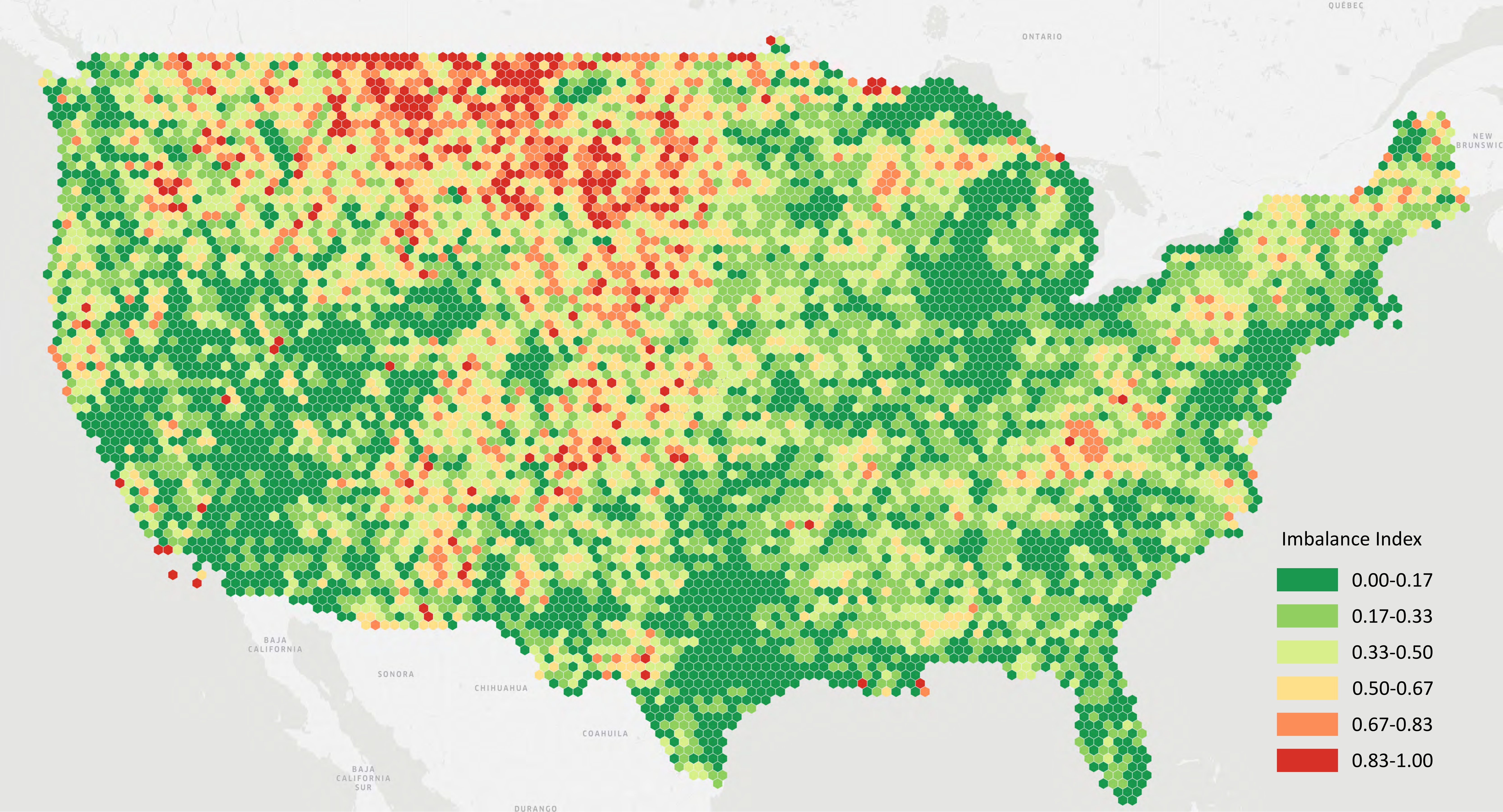}
         \caption{}
         \label{usa_grid}
     \end{subfigure}
     \hfill
     \begin{subfigure}[b]{0.45\textwidth}
         \centering
         \includegraphics[width=\textwidth]{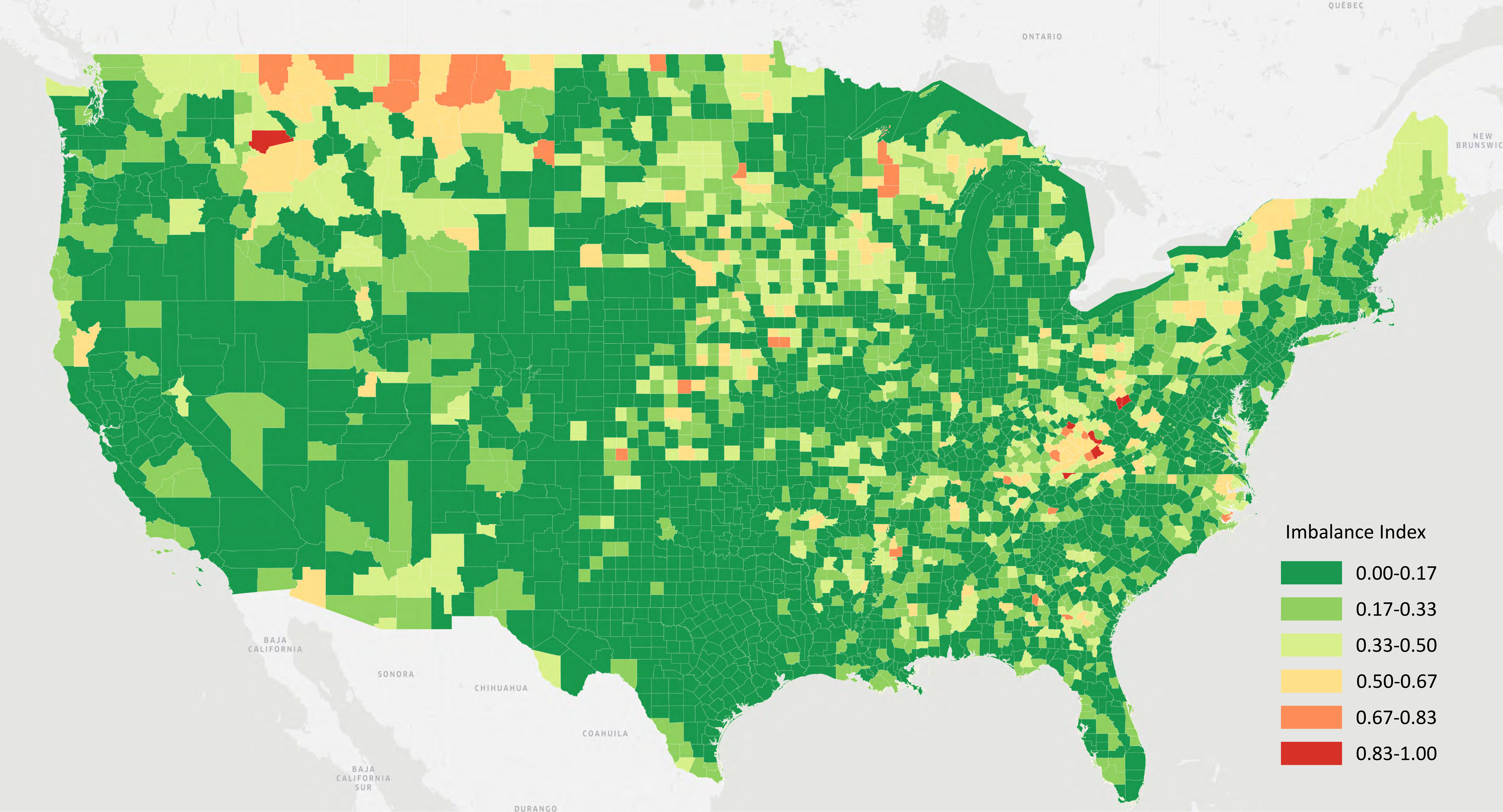}
         \caption{}
         \label{usa_division}
     \end{subfigure}
     
     \begin{subfigure}[b]{0.45\textwidth}
         \centering
         \includegraphics[width=\textwidth]{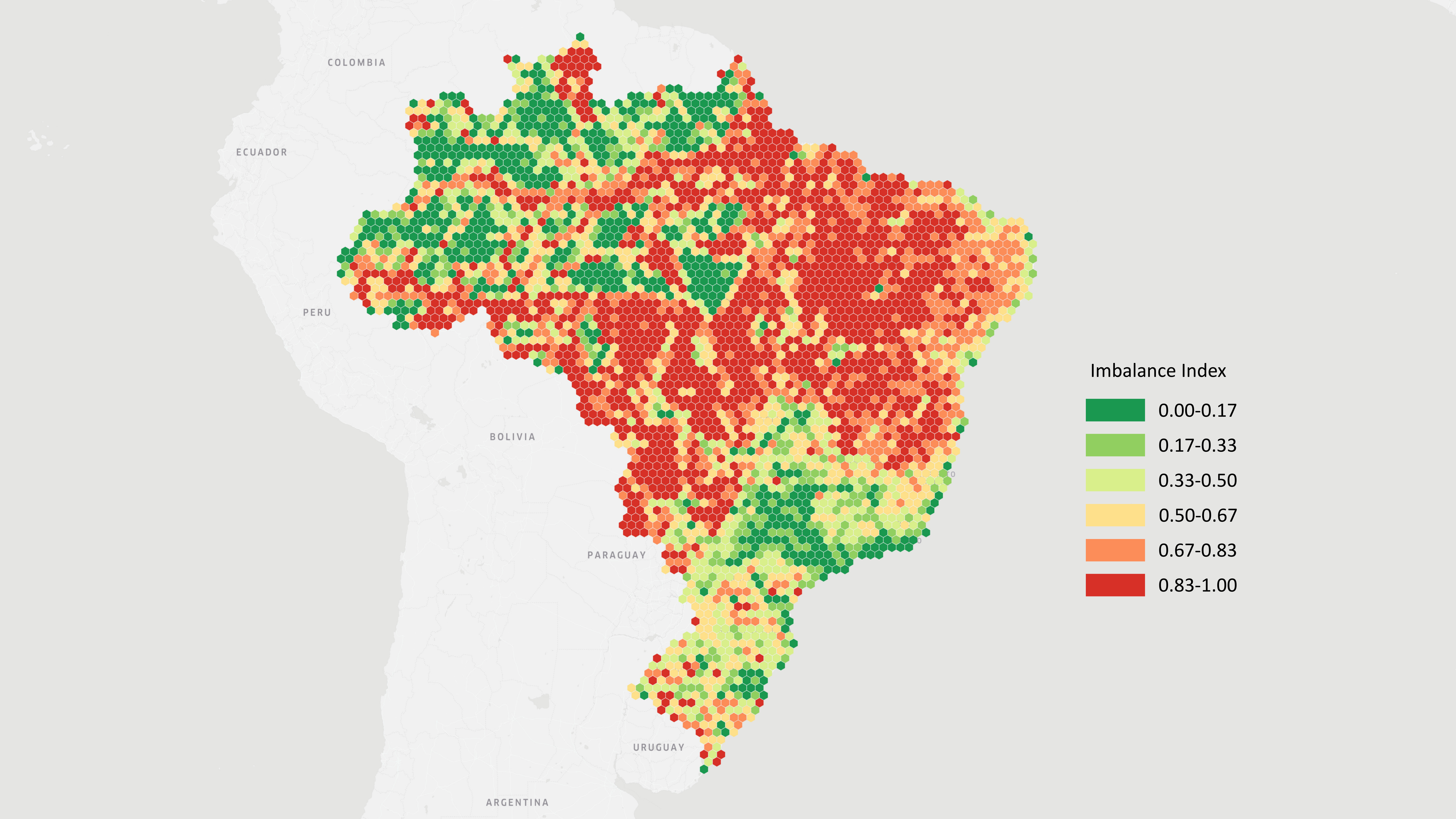}
         \caption{}
         \label{bra_division}
     \end{subfigure}
     \hfill
     \begin{subfigure}[b]{0.45\textwidth}
         \centering
         \includegraphics[width=\textwidth]{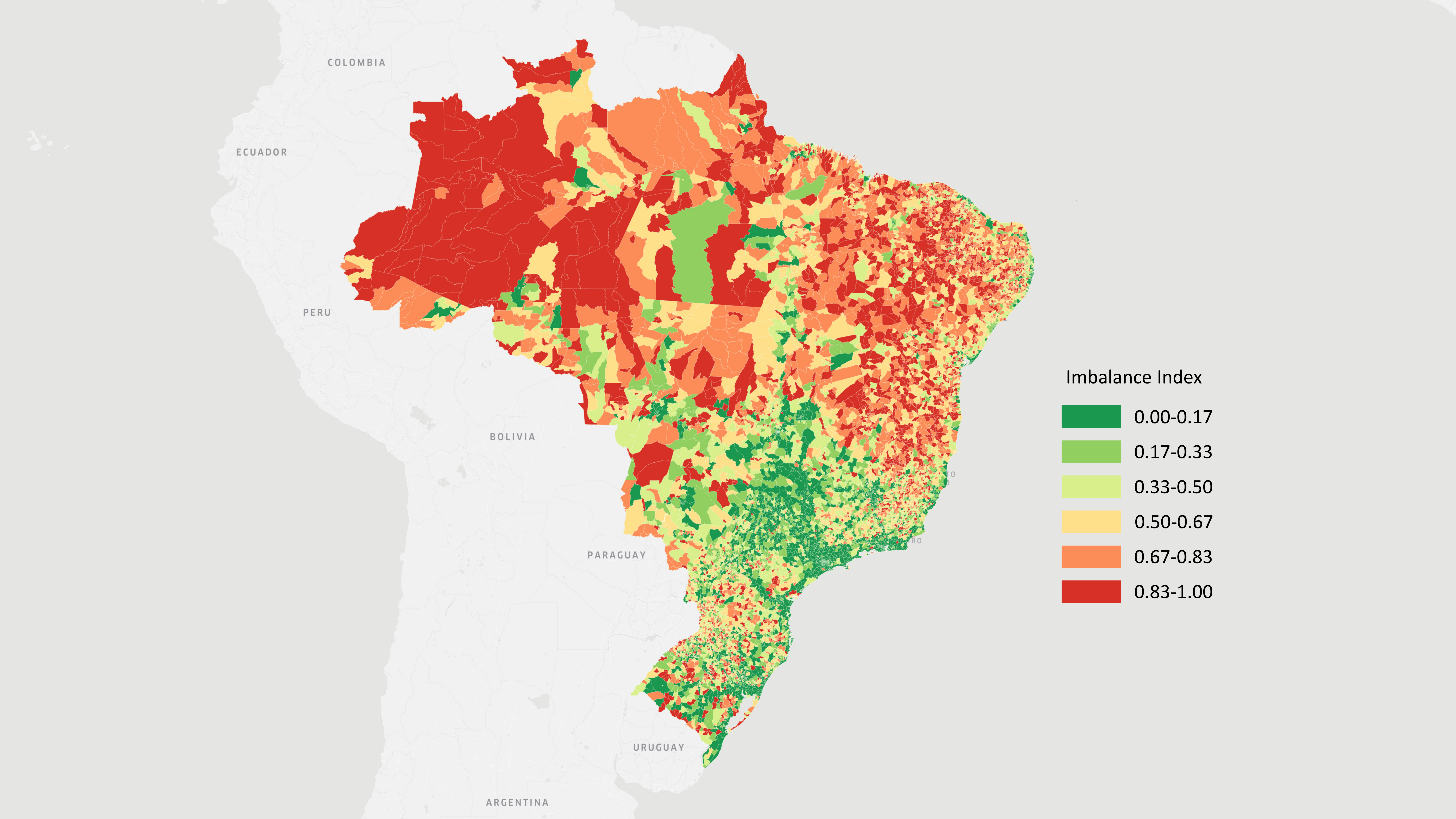}
         \caption{}
         \label{bra_division}
     \end{subfigure}
     
     \begin{subfigure}[b]{0.45\textwidth}
         \centering
         \includegraphics[width=\textwidth]{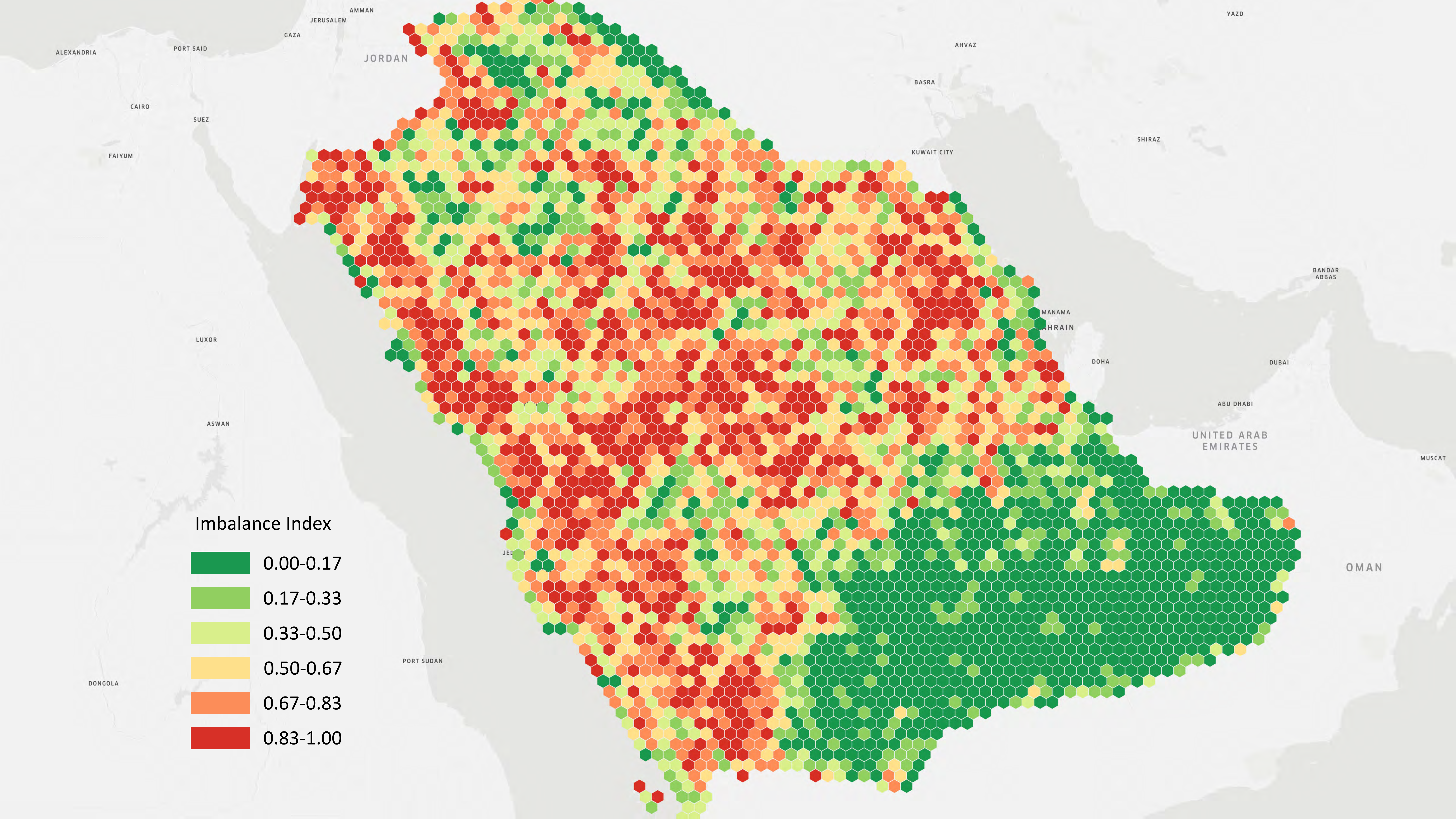}
         \caption{}
         \label{sau_division}
     \end{subfigure}
     \hfill
     \begin{subfigure}[b]{0.45\textwidth}
         \centering
         \includegraphics[width=\textwidth]{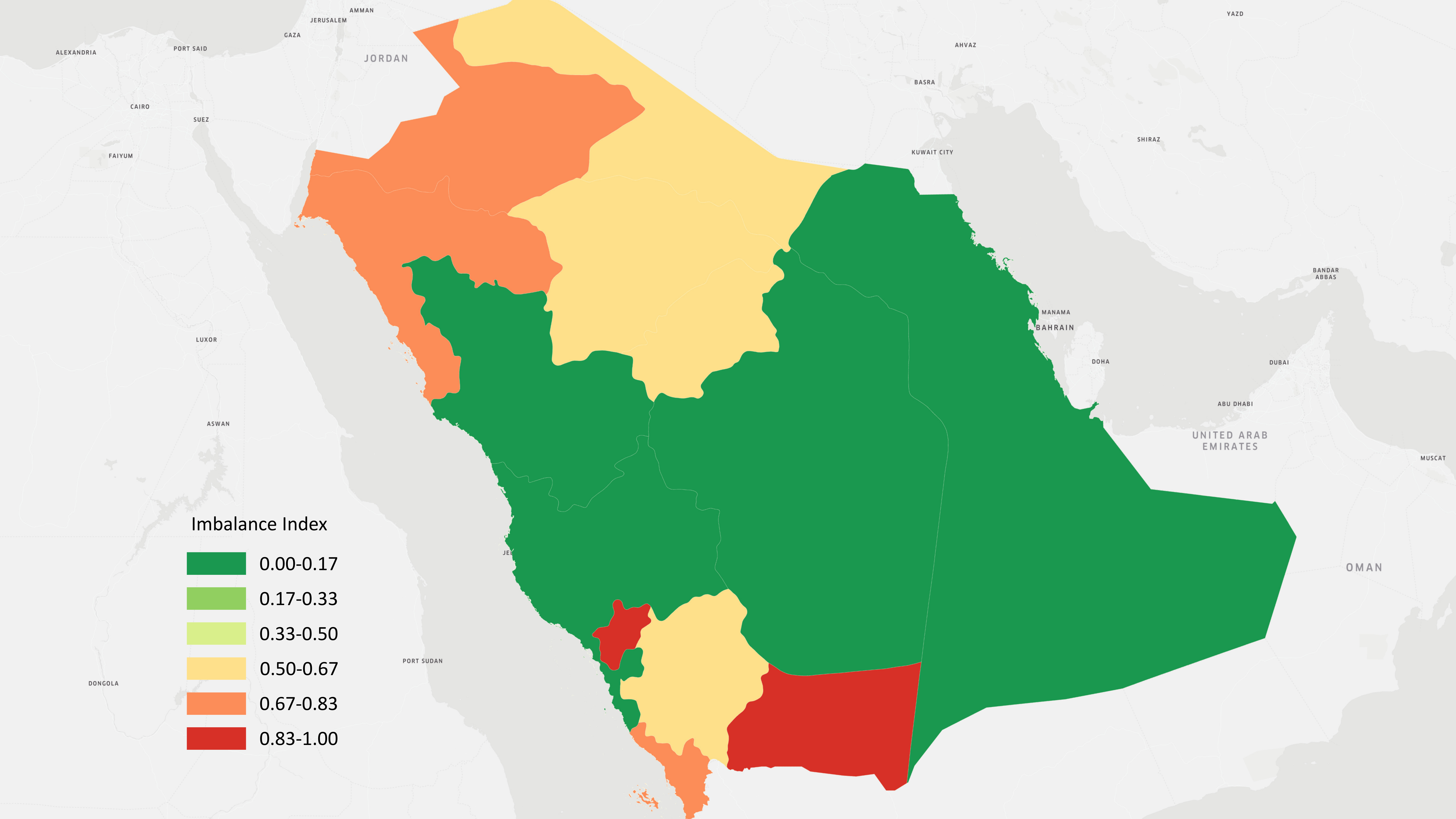}
         \caption{}
         \label{sau_division}
     \end{subfigure}
     
     \begin{subfigure}[b]{0.45\textwidth}
         \centering
         \includegraphics[width=\textwidth]{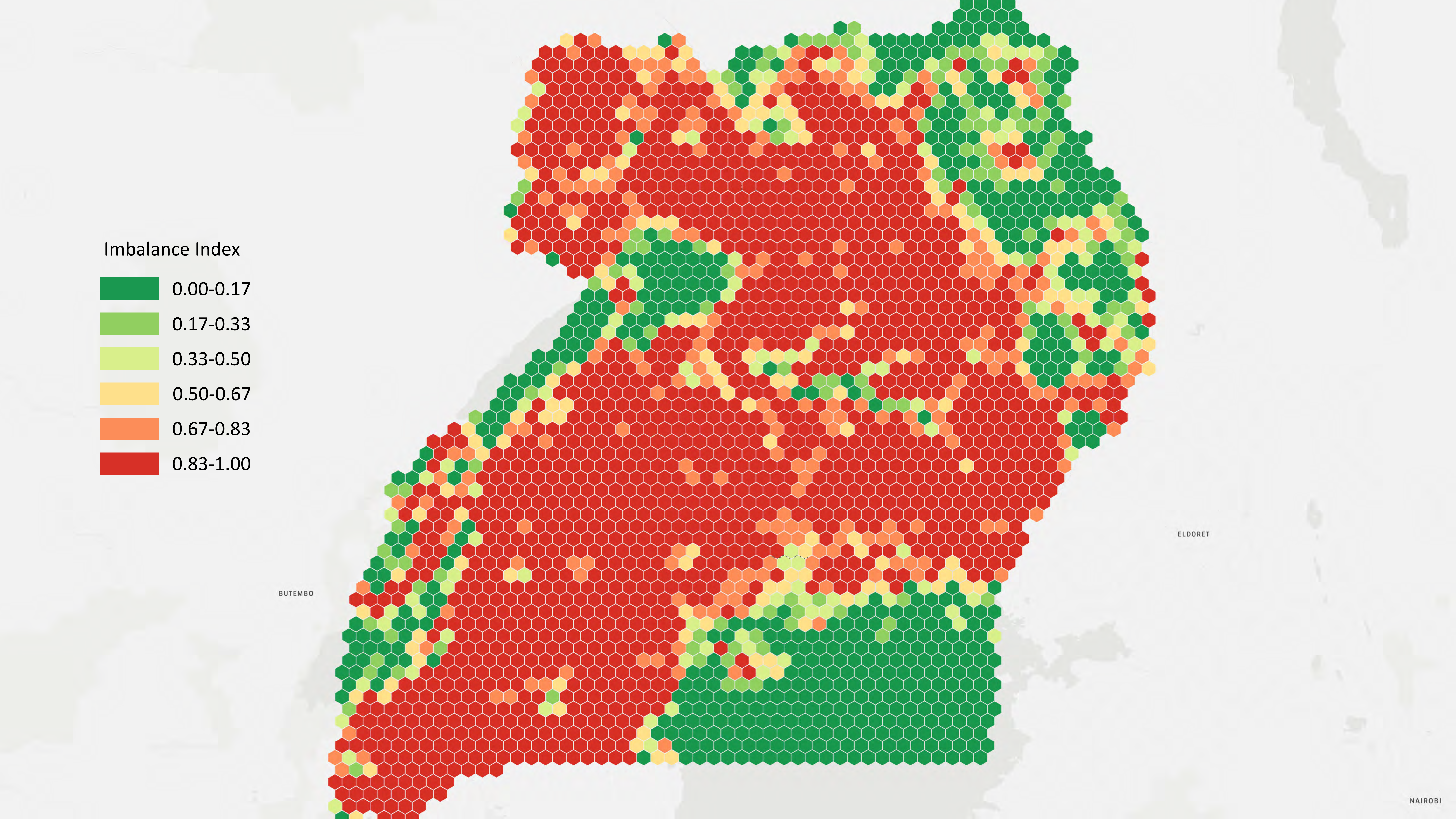}
         \caption{}
         \label{uga_division}
     \end{subfigure}
     \hfill
     \begin{subfigure}[b]{0.45\textwidth}
         \centering
         \includegraphics[width=\textwidth]{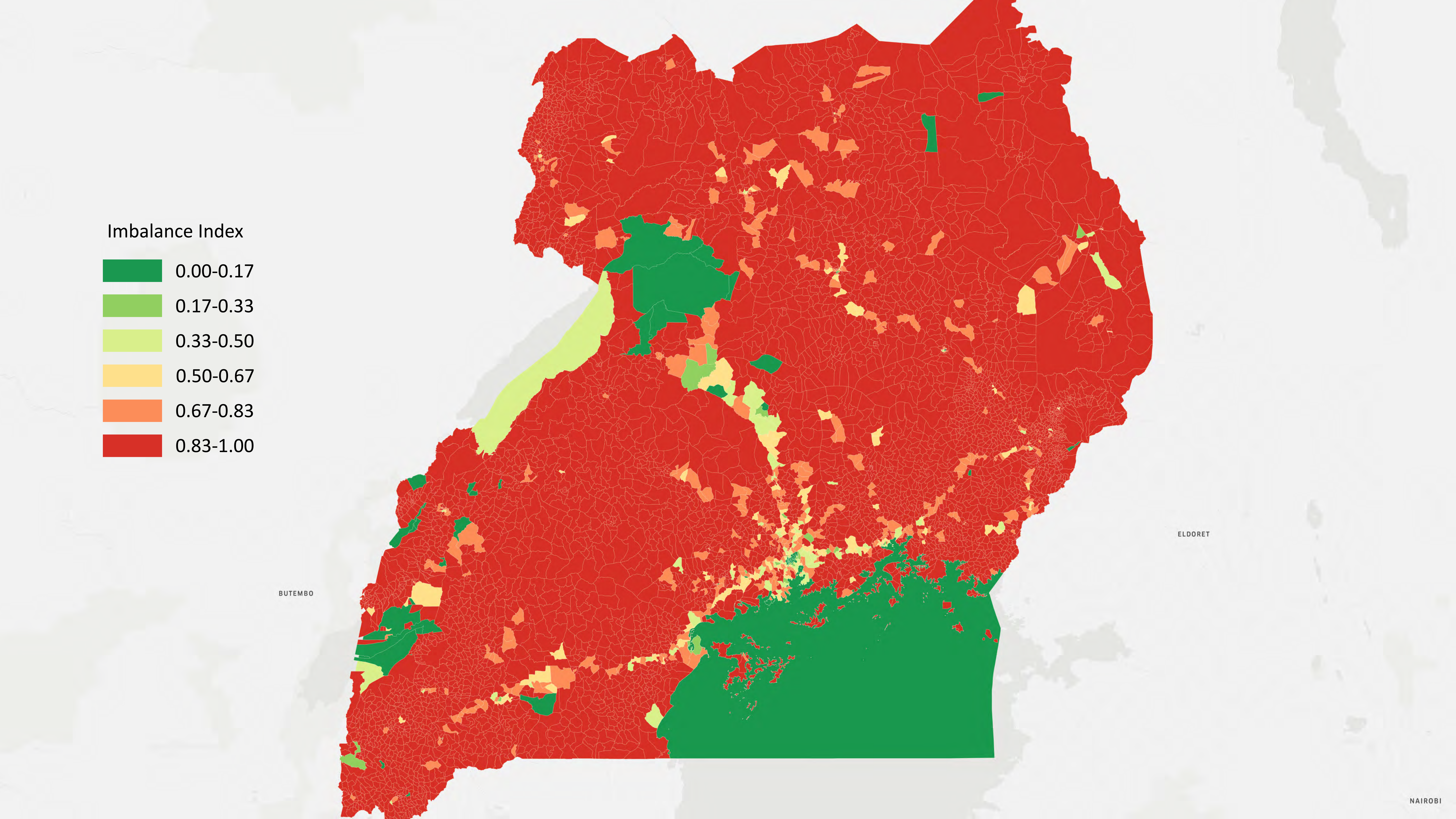}
         \caption{}
         \label{uga_division}
     \end{subfigure}
     
     \caption{Telecommunication service imbalance index visualization for the United States (Alaska and Hawaii are excluded for simplicity), Brazil, Saudi Arabia, and Uganda. Left: Grid-wise imbalance index visualization. Right: Administrative division-wise imbalance index visualization.}
        \label{index_first}
\end{figure}

\begin{figure}
     \centering
     \begin{subfigure}[b]{0.45\textwidth}
         \centering
         \includegraphics[width=\textwidth]{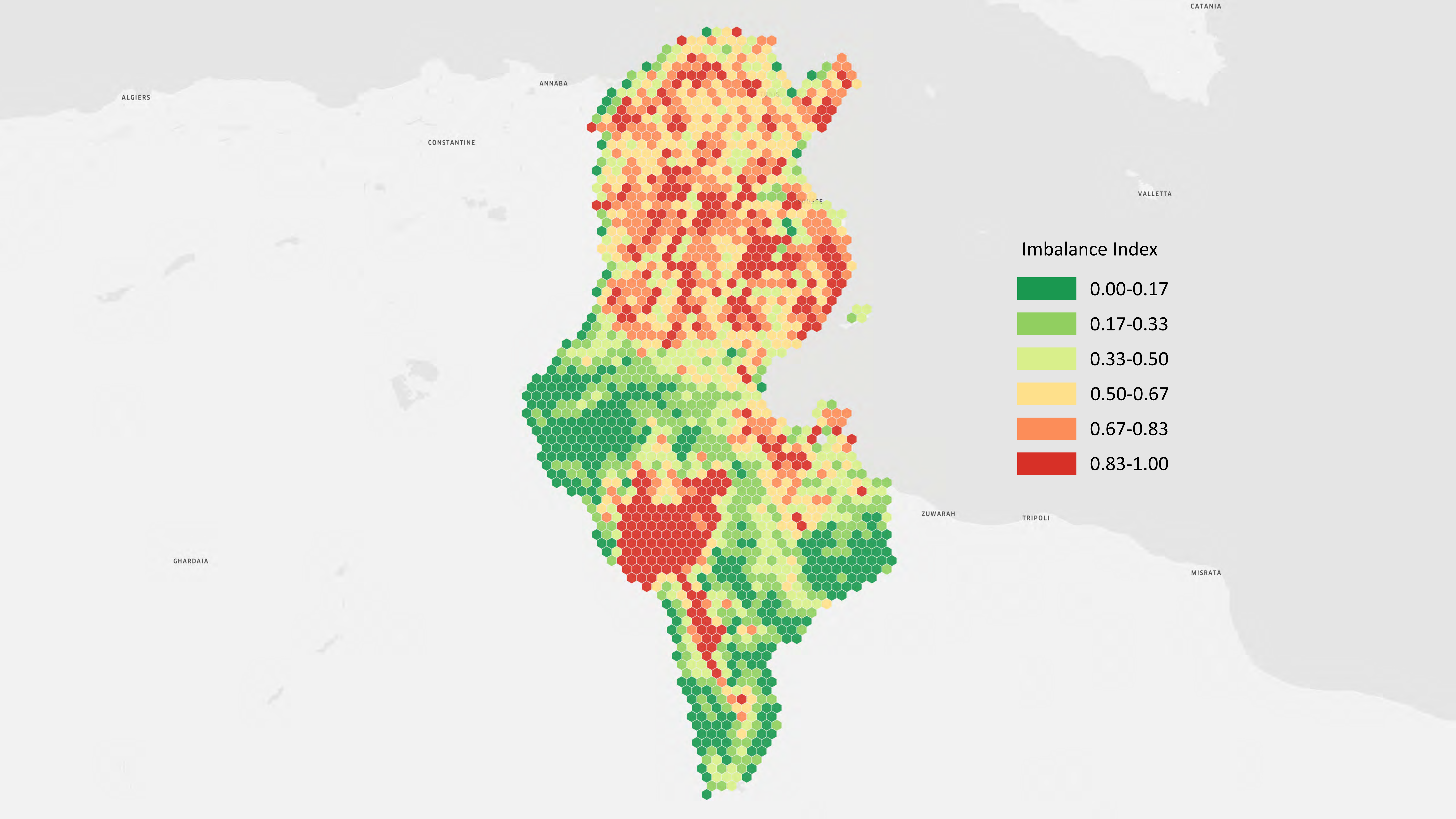}
         \caption{}
         \label{tun_grid}
     \end{subfigure}
     \hfill
     \begin{subfigure}[b]{0.45\textwidth}
         \centering
         \includegraphics[width=\textwidth]{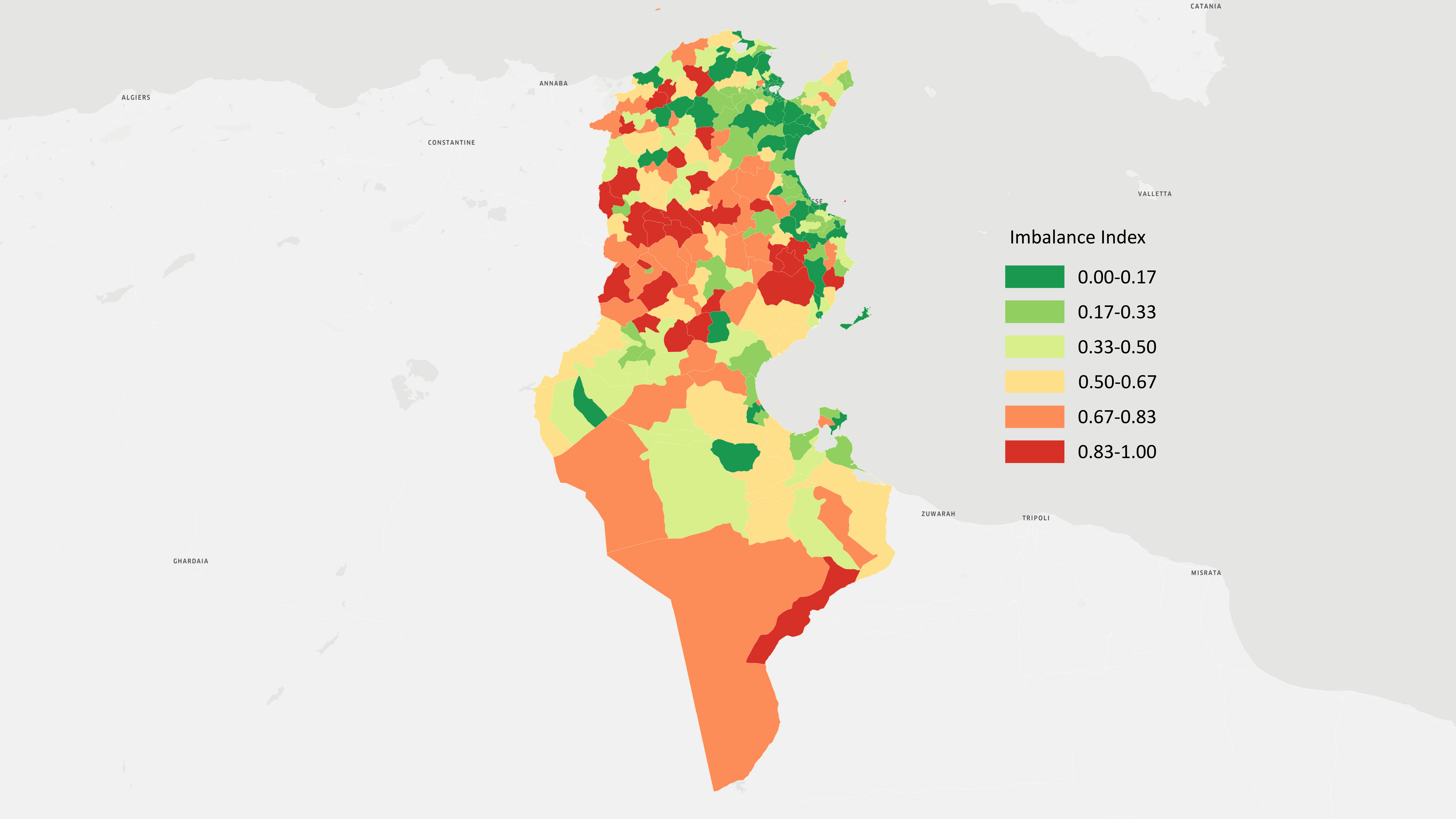}
         \caption{}
         \label{tun_division}
     \end{subfigure}
     
     \begin{subfigure}[b]{0.45\textwidth}
         \centering
         \includegraphics[width=\textwidth]{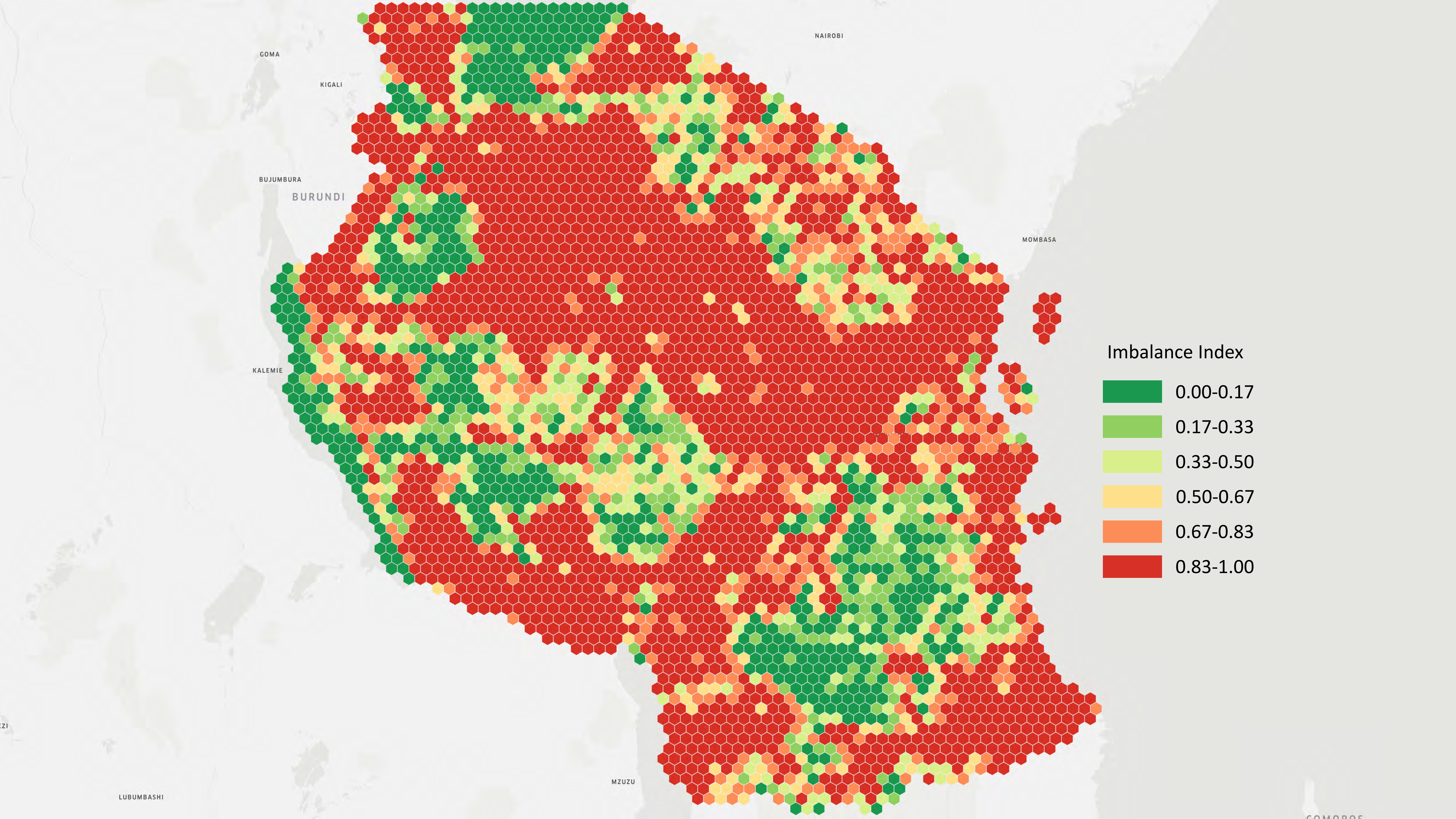}
         \caption{}
         \label{tza_division}
     \end{subfigure}
     \hfill
     \begin{subfigure}[b]{0.45\textwidth}
         \centering
         \includegraphics[width=\textwidth]{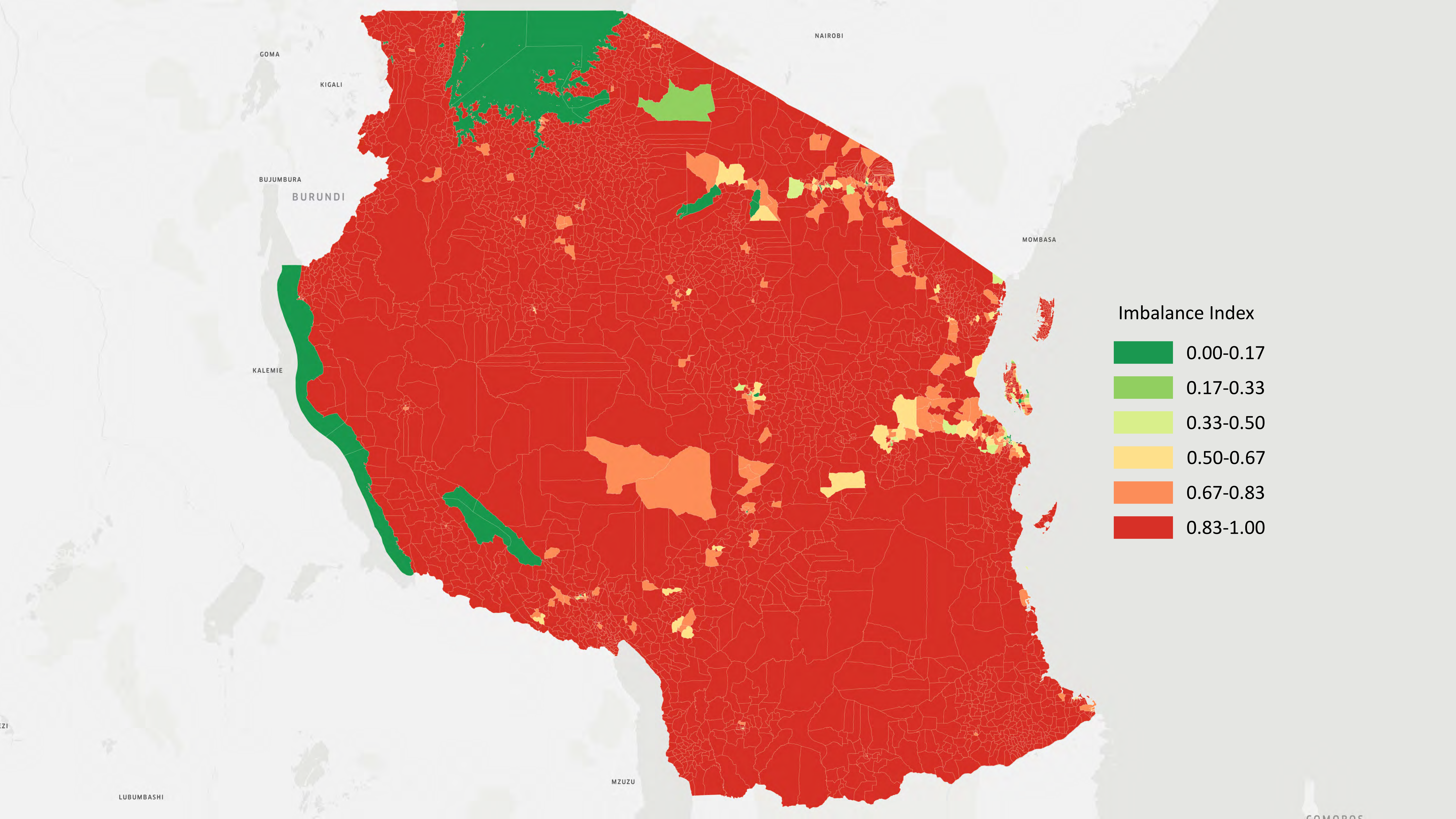}
         \caption{}
         \label{tza_division}
     \end{subfigure}
     
     \begin{subfigure}[b]{0.45\textwidth}
         \centering
         \includegraphics[width=\textwidth]{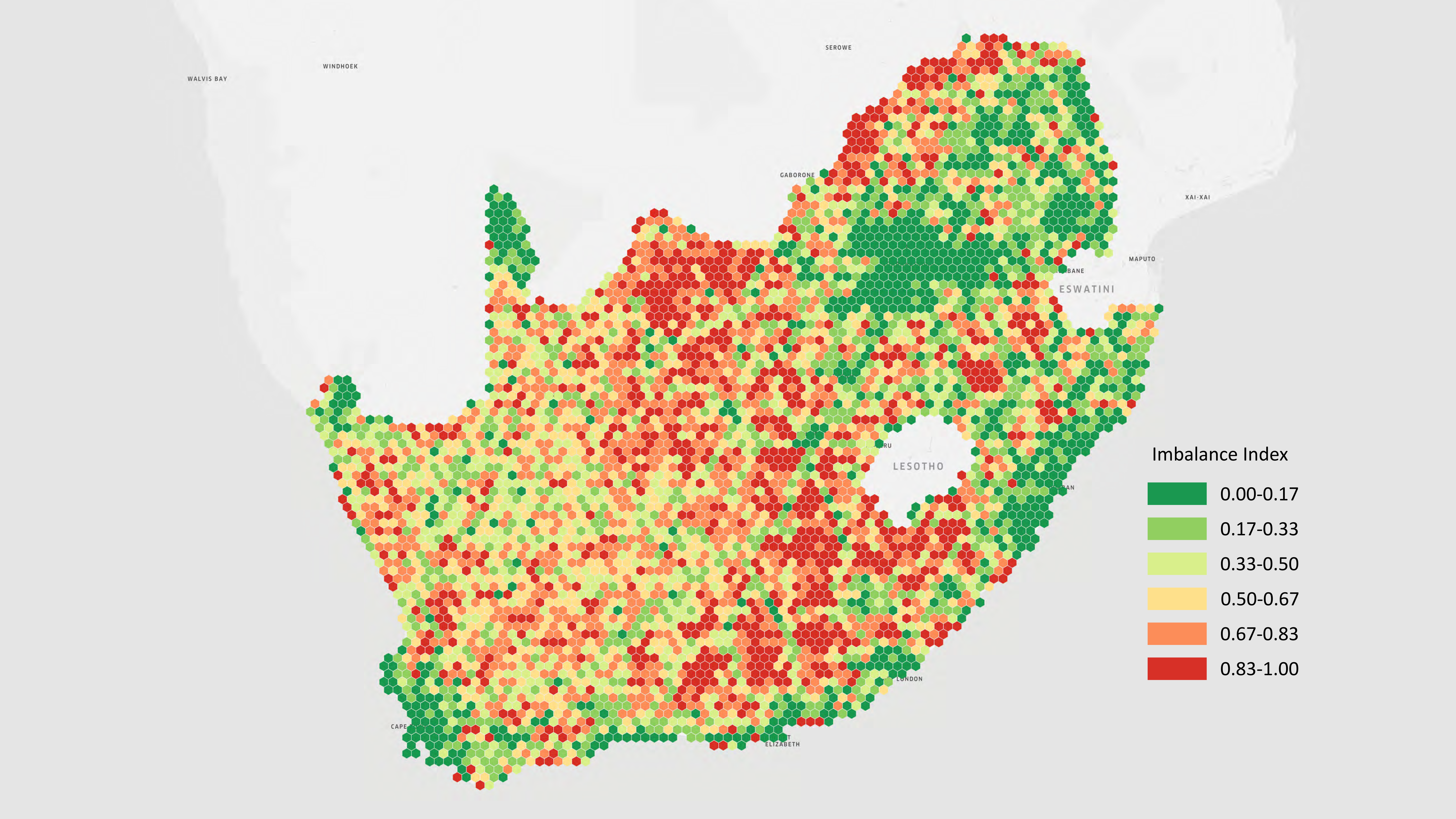}
         \caption{}
         \label{zaf_division}
     \end{subfigure}
     \hfill
     \begin{subfigure}[b]{0.45\textwidth}
         \centering
         \includegraphics[width=\textwidth]{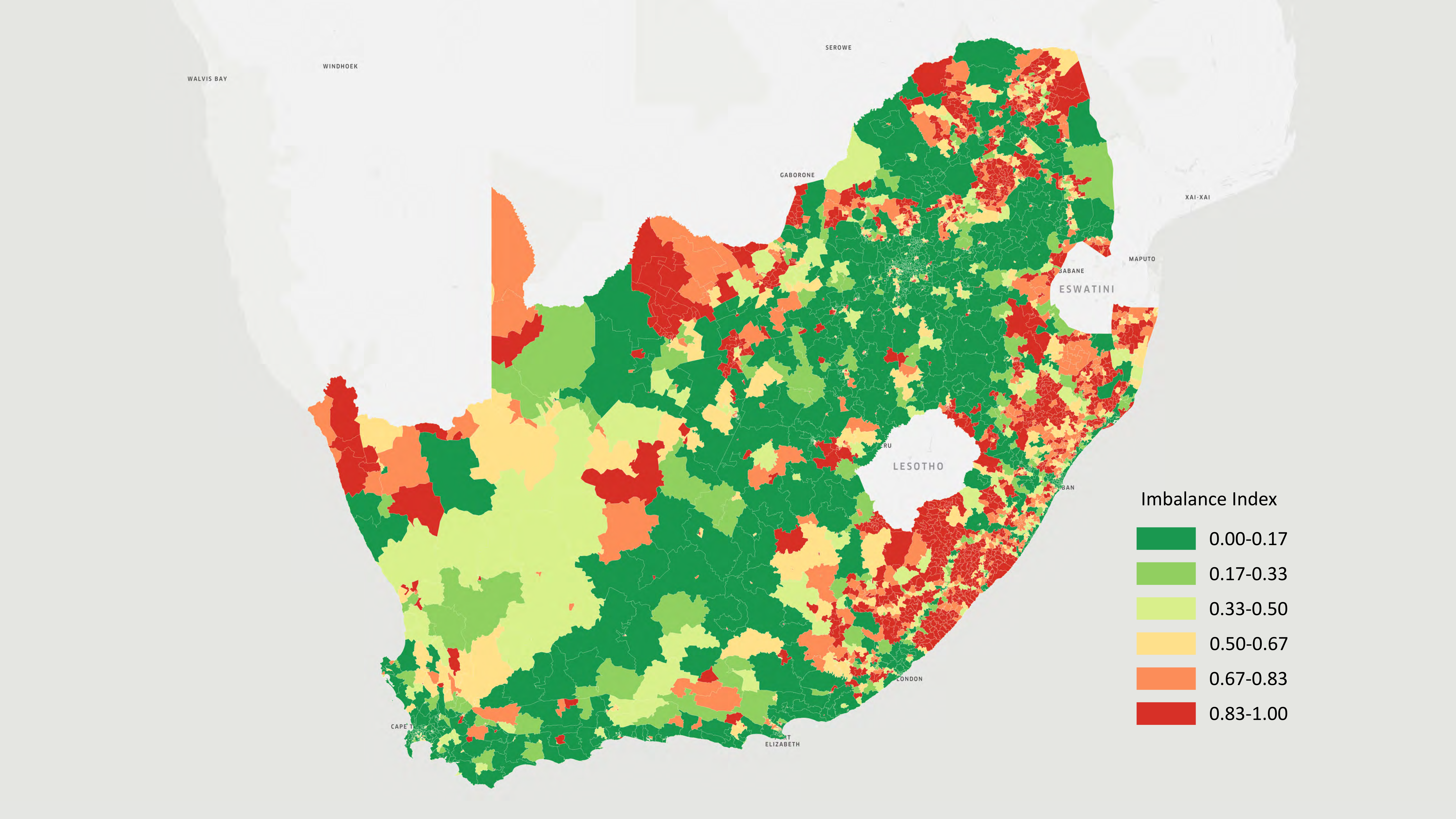}
         \caption{}
         \label{zaf_division}
     \end{subfigure}
     
     \begin{subfigure}[b]{0.45\textwidth}
         \centering
         \includegraphics[width=\textwidth]{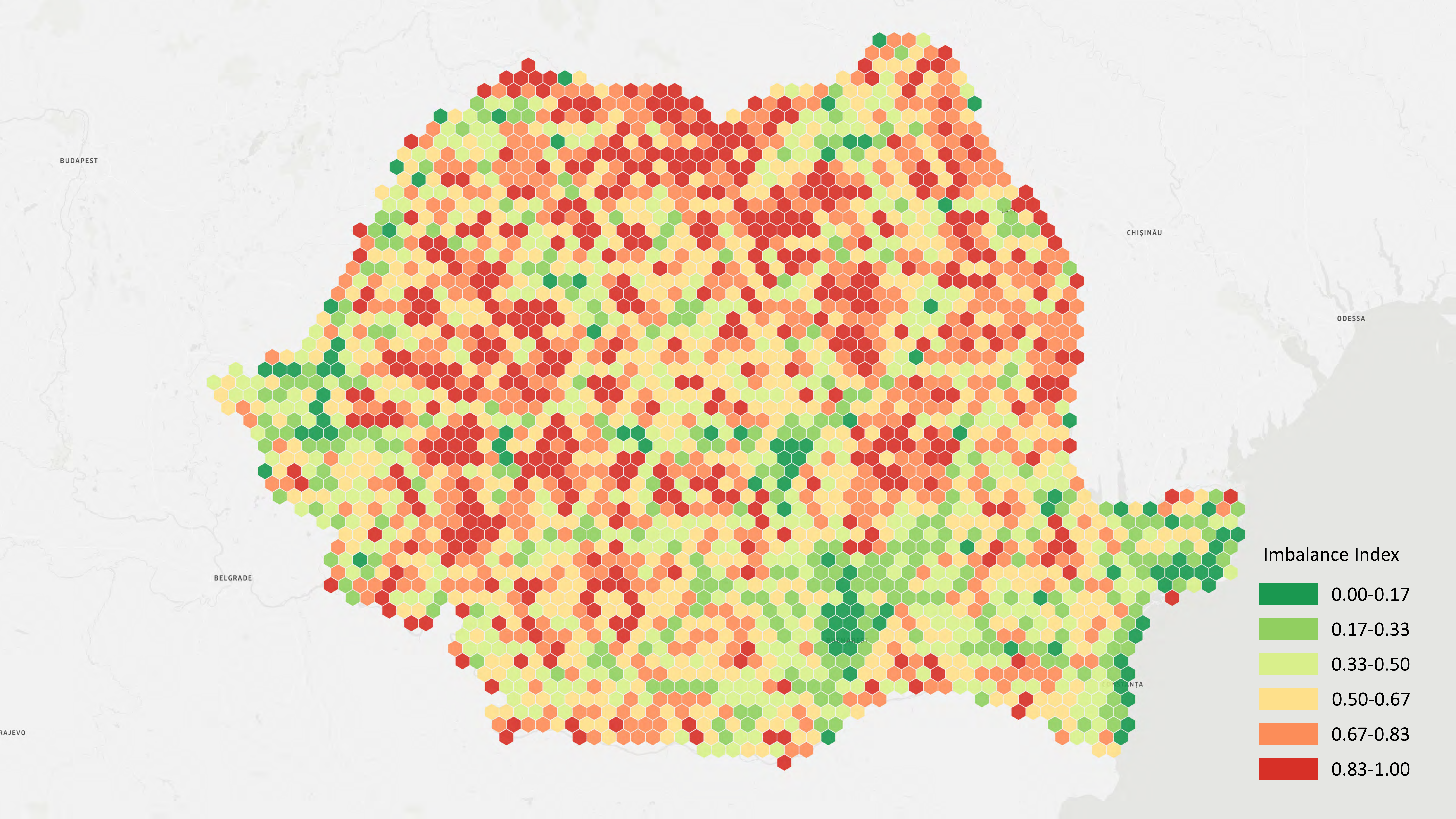}
         \caption{}
         \label{rou_division}
     \end{subfigure}
     \hfill
     \begin{subfigure}[b]{0.45\textwidth}
         \centering
         \includegraphics[width=\textwidth]{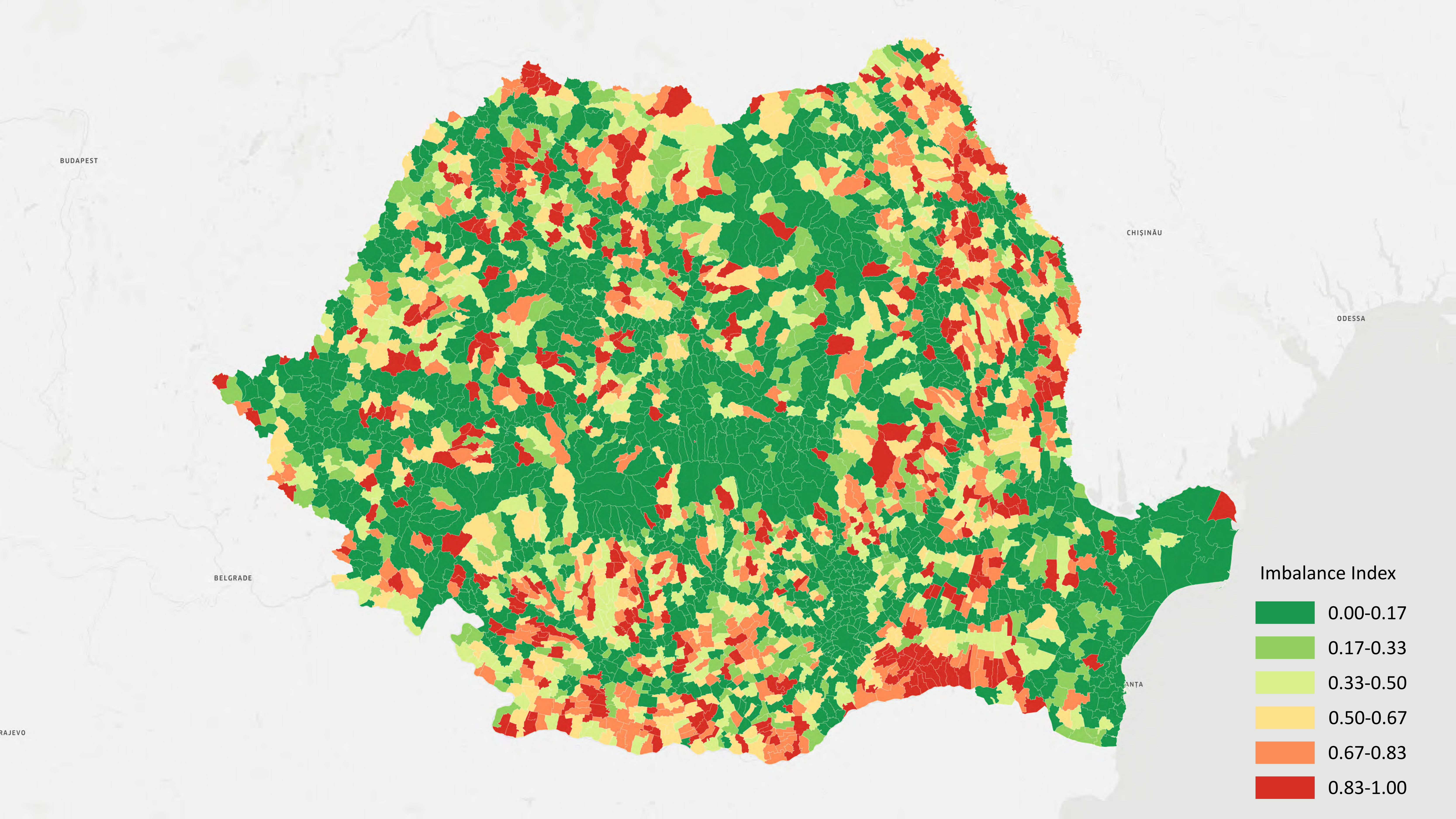}
         \caption{}
         \label{rou_division}
     \end{subfigure}
     
     \caption{Telecommunication service imbalance index visualization for Tunisia, Tanzania, South Africa, and Romania. Left: Grid-wise imbalance index visualization. Right: Administrative division-wise imbalance index visualization.}
        \label{index_second}
\end{figure}

\begin{figure}
     \centering
     \begin{subfigure}[b]{0.45\textwidth}
         \centering
         \includegraphics[width=\textwidth]{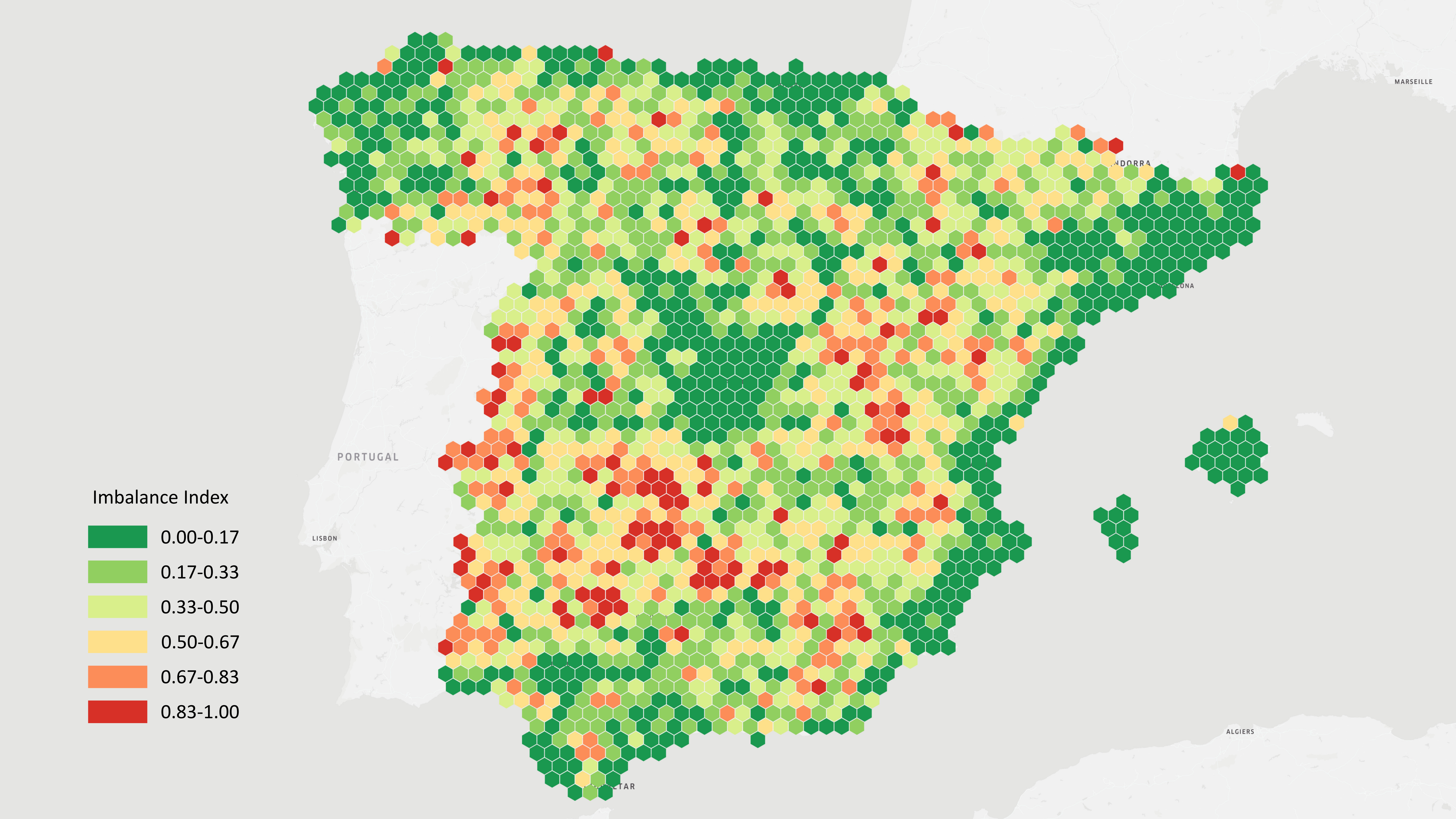}
         \caption{}
         \label{esp_grid}
     \end{subfigure}
     \hfill
     \begin{subfigure}[b]{0.45\textwidth}
         \centering
         \includegraphics[width=\textwidth]{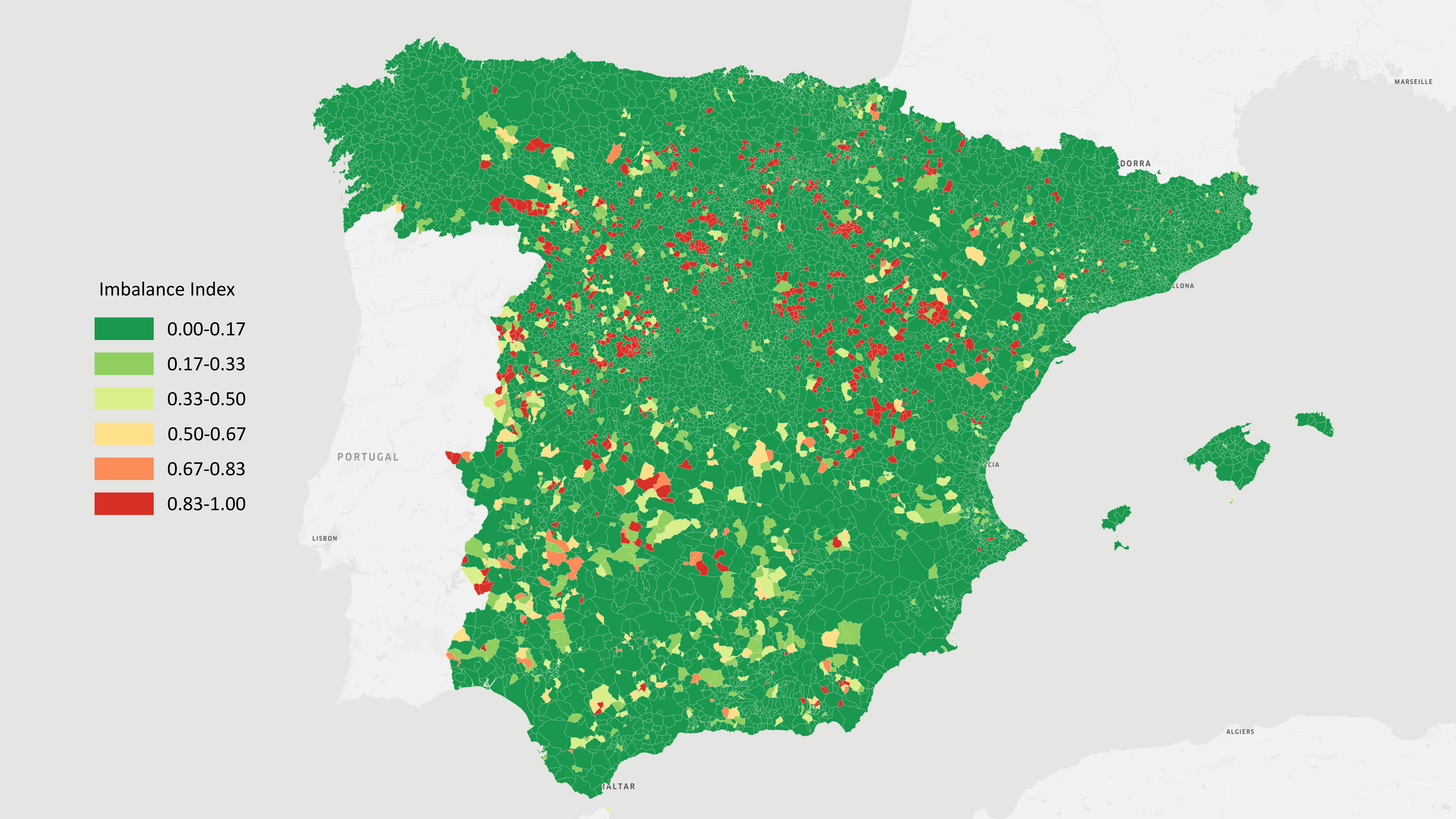}
         \caption{}
         \label{esp_division}
     \end{subfigure}
     
     \begin{subfigure}[b]{0.45\textwidth}
         \centering
         \includegraphics[width=\textwidth]{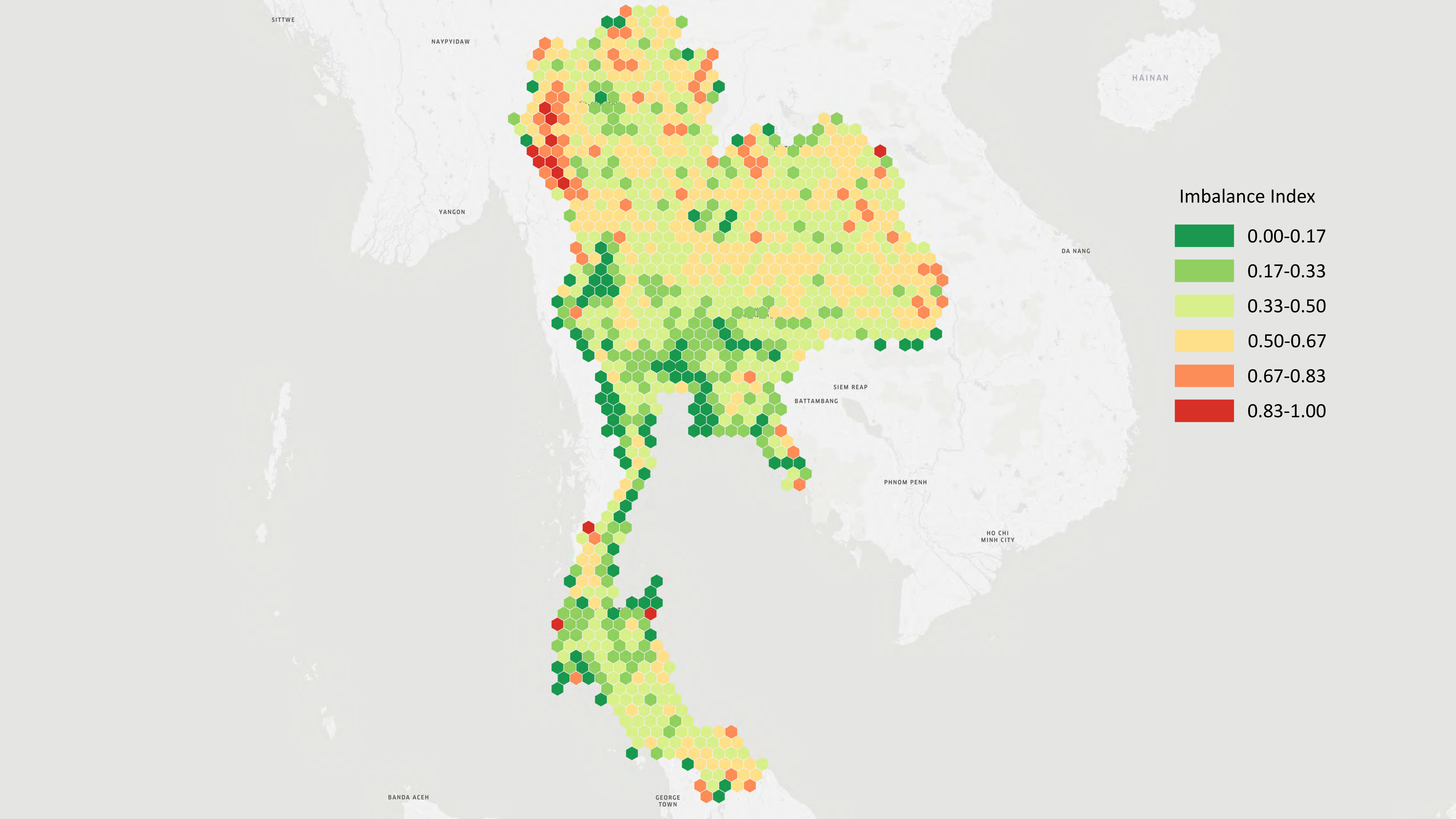}
         \caption{}
         \label{tha_division}
     \end{subfigure}
     \hfill
     \begin{subfigure}[b]{0.45\textwidth}
         \centering
         \includegraphics[width=\textwidth]{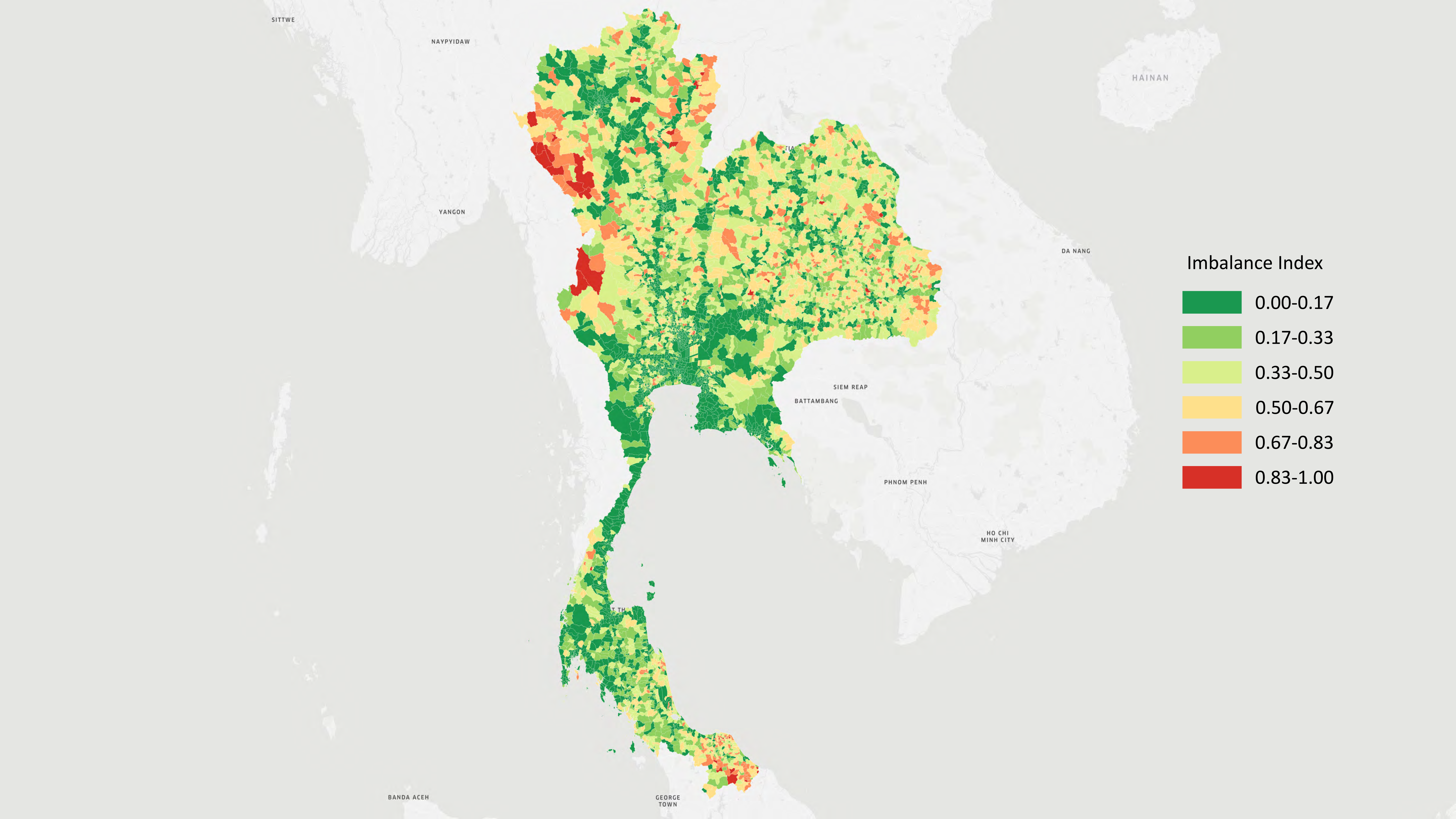}
         \caption{}
         \label{tha_division}
     \end{subfigure}
     \caption{Telecommunication service imbalance index visualization for Spain and Thailand. Left: Grid-wise imbalance index visualization. Right: Administrative division-wise imbalance index visualization.}
        \label{index_third}
\end{figure}

Finally, considering that the number of deployed BSs must be an integer, we can approximate the final solution of the number of deployed BSs within the $m$th granule as $B_m^\#=\lfloor B_m^*\rfloor$, where $\lfloor \cdot\rfloor$ is the floor function, returning the nearest integer equal to or smaller than the argument. Note that, because of the manipulation by the floor function, the number of placed BSs might be smaller than the total number of available BSs, i.e., $\Delta=B_{\max}-\underset{m\in\mathcal{R}(M)}{\sum}B_m^*>0$. Also, when $B_{\max}$ is small, it is also possible to have $\Delta=B_{\max}$, resulting in an invalid placement scheme. Therefore, in order to fully exploit $B_{\max}$ BSs, we propose the following auxiliary placement policies:
\begin{itemize}
\item When $\Delta>1$, we replace $B_{\max}$ by $\Delta$ in (\ref{opteq}) and repeat the placement optimization until $\Delta=1$ or $\Delta=0$;
\item When $\Delta=1$, we place the last remaining BS to the granule corresponding to the highest unserved population;
\item When $\Delta=B_{\max}$, we place one BS to the granule corresponding to the highest unserved population and then replace $B_{\max}$ by $B_{\max}-1$ in (\ref{opteq}) and repeat the placement optimization until $\Delta=1$ or $\Delta=0$;
\end{itemize}

\section{Experimental Results and Discussion}\label{erd}
This section is devoted to the visualization and validation of the proposed telecommunication service imbalance index, as well as the verification of the proposed BS placement strategy.

\subsection{Experiment Settings}
To obtain the imbalance index of a specific geographical segment, three parameters are involved, i.e., population $P$, the number of BS $B$, and the numbers of users that can be supported by existing BSs $\boldsymbol{\rho}(B)$. The population data comes from Facebook's High Resolution Population Density Maps and Demographic Estimates that can be found in \cite{fcl}. This dataset includes the estimated number of people living within 30-meter grid tiles in nearly every country around the globe, which is extremely fine-grained. The worldwide BS dataset comes from the OpenCellid Project of Unwired Labs \cite{opencellid}. The OpenCellid dataset contains the estimated location of each BS all over the world. As for $\boldsymbol{\rho}(B)$, for simplicity, we assume that all BSs can support the same number of users, and in our experiment, we set $\{\rho_i=100\}_{i=1}^{B}$. 
There are another two hyper-parameters in (\ref{inequalityindexsinglegreq}), that is, $\alpha$ and $\beta$, and we normalize both in the following experiments without loss of generality.

\begin{figure*}[!ht]
\centering
\includegraphics[width=0.95\textwidth]{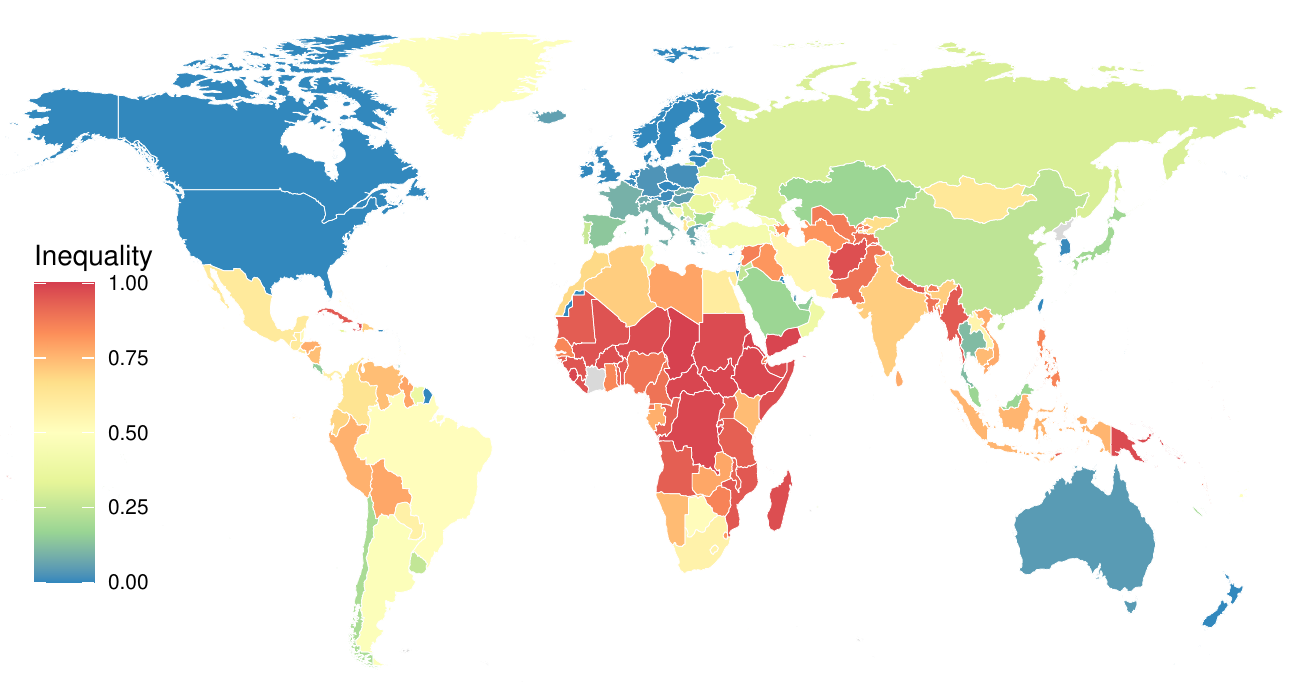}
\caption{Visualization of the telecommunication service imbalance for the whole world by the proposed imbalance index.}
\label{world_wide_inequality}
\end{figure*}

\subsection{Visualization of Telecommunication Service Imbalance}
Based on the above settings, we can calculate the telecommunication service imbalance index for each region of a country for illustration purposes. Taking Vietnam as an example, the calculated imbalance indexes (both in grid-wise and administrative division-wise) are displayed in Fig. \ref{index_vnm}. From this figure, one can clearly observe the imbalance index of each region across the whole country. Some regions, such as those near the capital Hanoi and the biggest city Ho Chi Minh City, are of low imbalance, which indicates satisfactory telecommunication services. Efforts of mobile network operators in improving the connectivity in these regions are superfluous. On the contrary, they should pay particular attention to those regions of high imbalance, since few infrastructure resources deployed would likely bring huge social and economic benefits. More countries' imbalance index visualization can be found in Fig. \ref{index_first}, Fig. \ref{index_second}, and Fig. \ref{index_third}. For simplicity, we omit the detailed explanation for these countries.

Beyond the region-wise service imbalance index, we go one step further and plot the country-wise service imbalance index across the globe in Fig. \ref{world_wide_inequality}. In this figure, each country's imbalance index is obtained by averaging all region's indexes within that country.
As reflected in this figure, we can tell that Europe and North America have lower telecommunication service imbalance indexes. South America and East Asia have moderate imbalance indexes. South Asia and Africa have relatively high imbalance indexes. These results are positively related to the levels of development of these countries and are of paramount importance for understanding global informatization and the digital divide.

\begin{figure}[!ht]
\centering
\includegraphics[width=0.6\textwidth]{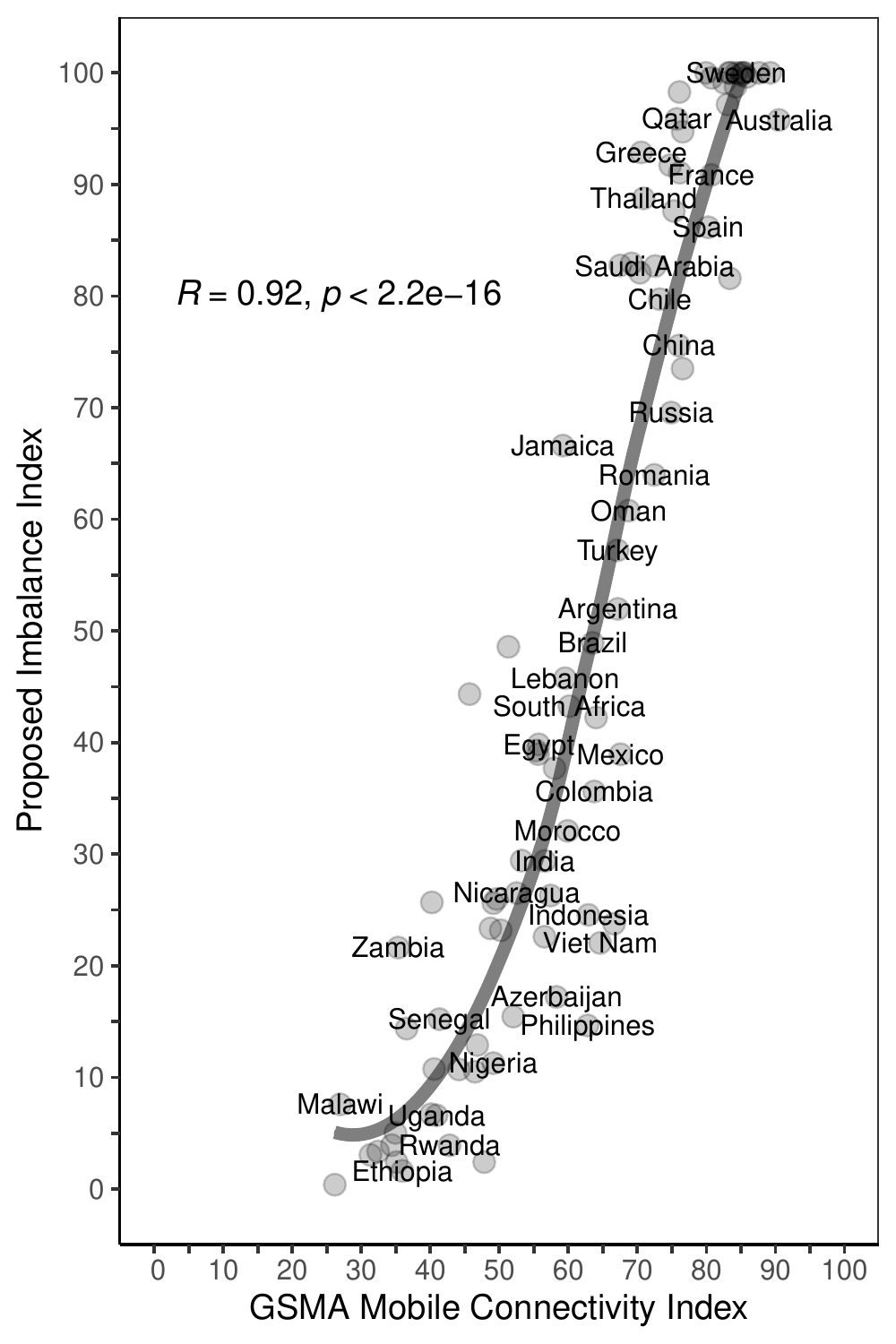}
\caption{Correlation analysis between the GSMA mobile connectivity index and the proposed telecommunication service imbalance index.}
\label{corr_gsma}
\end{figure}

\subsection{Consistency with Coarse-Grained Measures}

We validate the effectiveness of our imbalance index by comparing it with a coarse-grained measure, i.e., the GSMA mobile connectivity index \cite{GSMA_2020}.  We calculate the imbalance indexes for ninety countries and analyze the correlation between our index and the GSMA index. A scatter plot and the corresponding Pearson correlation coefficient are displayed in Fig. \ref{corr_gsma}. Note that, in this figure, we rescale our index by $(1.0 - \lambda) * 100$ to align it with the GSMA index.
We can observe from Fig. \ref{corr_gsma} that the relationship between the proposed index and the GSMA index is almost linear. The derived Pearson correlation coefficient $R$ is 0.92, indicating a high correlation level existing between the proposed index and the GSMA index. The achieved statistical $p$-value is $2.2e^{-16}$, which is sufficiently small to reject the null hypothesis (without correlation), and therefore the alternative hypothesis (with true correlation) should be accepted.
Through the empirical data and the corresponding analysis, we can summarize that our index is consistent with coarse-grained measures, indicating the effectiveness of our proposed index for quantifying the telecommunication service imbalance. However, compared with the GSMA index that provides only country-wise statistics, our index is fine-grained and can provide imbalance statistics for a small geographical segment.


\subsection{Verification for the Base Station Placement Strategy}
As there are a number of practical constraints that are dropped in the formulated problem in (\ref{opteq}) for simplifying our discussion and reveal the essence of the problem, we might temporarily study the BS placement problem for a virtual case through simulations. In particular, we assume a $100~\mathrm{km}\times 100~\mathrm{km}$ square region where the population is distributed abiding the Mat\'{e}rn cluster process\footnote{The Mat\'{e}rn cluster process is used to characterize  clustering behaviors and patterns of points in space and has been verified to be an effective model for many practical applications \cite{Tang2017}.  Note that, for simplicity, we neglect the boundary effect and discard the generated points beyond the $100~\mathrm{km}\times 100~\mathrm{km}$ square window.}. The parent and off-spring densities are $\theta_p=5\times10^{-9}~\mathrm{person/m}^2$ and $\theta_o=\times10^{-6}~\mathrm{person/m}^2$, and the cluster radius $r=10~\mathrm{km}$.

Without loss of generality, we consider several different gridding schemes with $20~\mathrm{km}\times 20~\mathrm{km}$ grid (25 granules) and $50~\mathrm{km}\times 50~\mathrm{km}$ grid (4 granules). We adopt the following two placement strategies as comparison benchmarks:
\begin{itemize}
\item Lower-bounding strategy: Release the constraint that $B_m$ must be an integer;
\item Na\"{i}ve placement strategy: Place BSs in proportion to populations in granules.
\end{itemize}

\begin{figure}[!t]
\centering
\includegraphics[width=0.95\textwidth]{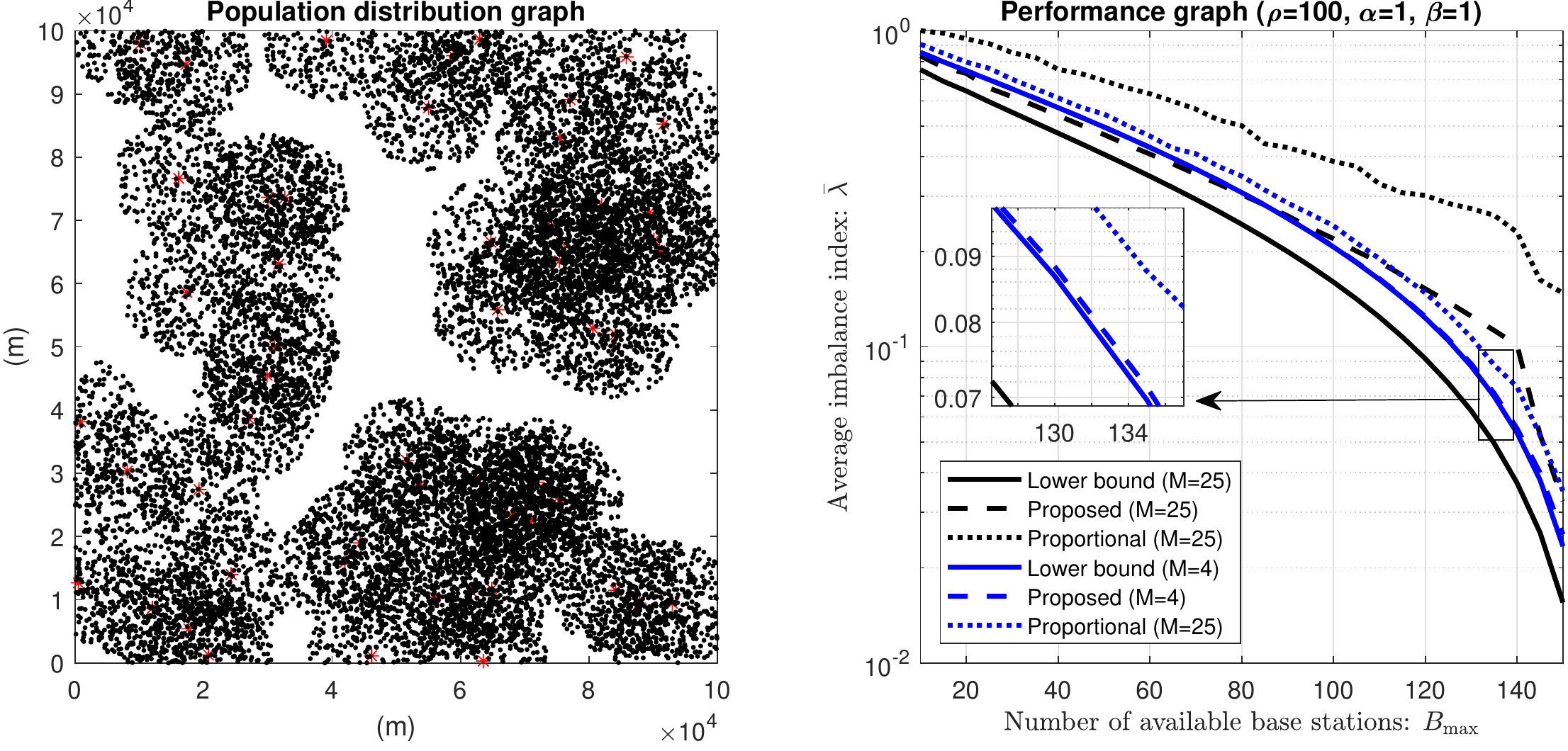}
\caption{(a) A realization of the Mat\'{e}rn cluster process, where red stars represent parent points and black dots represent offspring points, given $\theta_p=5\times10^{-9}~\mathrm{person/m}^2$, $\theta_o=\times10^{-6}~\mathrm{person/m}^2$, and $r=10~\mathrm{km}$; (b) Average imbalance indexes vs. the number of available BSs for placement.}
\label{matern}
\end{figure}

Based on the aforementioned settings, a realization of the Mat\'{e}rn cluster process and the corresponding simulation results regarding the average imbalance index are presented in Fig. \ref{matern}. From this figure, it can be shown that the proposed placement strategy is close to the lower-bounding strategy when $B_{\max}$ goes large, which verifies its optimality to the placement problem formulated in (\ref{opteq}) for large $B_{\max}$. Also, when $B_{\max}$ is sufficiently large, the proposed placement strategy is also better than the Na\"{i}ve placement strategy. In addition, by comparing the cases with $M=25$ and $M=4$, we can also observe that increasing $M=25$ will lead to a lower average imbalance index given the same population distribution and available telecommunication resources. This reduction on average imbalance index is aligned with our expectation and reflects the advantage of delicacy management relying on fine-grained measures.

\section{Conclusion}\label{c}
In this letter, we proposed a fine-grained index on measuring telecommunication service imbalance. Through this index, we established a concrete mathematical relationship among connectivity imbalance, demographics, and telecommunication infrastructures. The proposed index is given in a general form that can be computed easily and configured flexibly to meet the needs of different policy makers or mobile network operators.
Based on the proposed imbalance index, we further proposed an infrastructure resource placement strategy to democratize the benefits of ICT.
We validated the effectiveness of the proposed imbalance index by showing its correlation with the GSMA mobile connectivity index that is a widely used coarse-grained index. Experimental results also confirmed the superiority of the proposed infrastructure resource placement strategy aiming to minimize the imbalance index.

\bibliographystyle{IEEEtran}
\bibliography{bib}

\end{document}